\documentclass[reqno]{amsart}

\usepackage[usenames,dvipsnames,svgnames,table]{xcolor}
\usepackage{amssymb,latexsym,amsfonts,amsmath}
\usepackage{graphicx}
\usepackage{eso-pic}
\usepackage{epsfig}
\usepackage{mathtools}
\usepackage{mdwlist}
\usepackage{refcount}
\usepackage{comment}
\usepackage{cite}
\usepackage[hidelinks]{hyperref}

\usepackage{keyval,trig}
\usepackage{amssymb,dsfont}
\usepackage{mathtools,mathrsfs}

\usepackage{amsthm}
\newtheorem{theorem}{Theorem}
\newtheorem{lemma}{Lemma}

\newtheorem{problem}{Problem}

\newtheorem{definition}{Definition}

\newtheorem{remark}{Remark}

\usepackage{graphicx}
\graphicspath{ {./images/}}
\usepackage{tikz}

\newcommand{\simF}{\mathcal{V}}
\newcommand{\simFfull}{\mathcal{V}(\mathbf{x}_1,\mathbf{x}_2)}
\def\x{\mathbf{x}}
\def\v{\mathbf{v}}
\def\w{\mathbf{w}}
\def\u{\mathbf{u}}
\def\d{\mathbf{d}}
\def\y{\mathbf{y}}
\def\Mbar{\bar{M}}

\def\reals{\mathbb{R}}

\newcommand{\LTLalways}{\Box}
\newcommand{\LTLeventually}{\Diamond}

\usepackage[T1]{fontenc}
\usepackage[utf8]{inputenc}

\newcommand{\word}{\omega}

\usepackage[many]{tcolorbox}
\usetikzlibrary{calc}
\tcbuselibrary{skins}

\newtcolorbox{resp}[1][]{%
	enhanced jigsaw,%
	colback=gray!5!white,%
	colframe=gray!80!black,%
	size=small,%
	boxrule=1pt,%
	halign title=flush center,%
	coltitle=black,%
	breakable,%
	drop shadow=black!50!white,%
	attach boxed title to top left={xshift=1cm,yshift=-\tcboxedtitleheight/2,yshifttext=-\tcboxedtitleheight/2},%
	minipage boxed title=3cm,%
	boxed title style={%
		colback=white,%
		size=fbox,%
		boxrule=1pt,%
		boxsep=2pt,%
		underlay={%
			\coordinate (dotA) at ($(interior.west) + (-0.5pt,0)$);
			\coordinate (dotB) at ($(interior.east) + (0.5pt,0)$);
			\begin{scope}[gray!80!black]
				\fill (dotA) circle (2pt);
				\fill (dotB) circle (2pt);
			\end{scope}
		}%
	},%
	#1%
}

\usepackage{ifthen,version}
\newboolean{arxiv-version}
\setboolean{arxiv-version}{true}
\ifthenelse{\boolean{arxiv-version}}{\includeversion{prnt-arxiv}}%
{\excludeversion{prnt-arxiv}}
\ifthenelse{\boolean{arxiv-version}}{\excludeversion{prnt-cready}}%
{\includeversion{prnt-cready}}

\usepackage{booktabs} % For formal tables
\usepackage{IEEEtrantools}
\usepackage{amssymb,latexsym,amsfonts,amsmath,amsthm}
\usepackage{thmtools, thm-restate}

\usepackage{mdframed,lipsum}

\usepackage{graphicx}

\usepackage{mathrsfs}

\usepackage{graphicx}
\usepackage{paralist}
\usepackage{dsfont}
\usepackage{eso-pic}
\usepackage{epsfig}
\usepackage{mathtools}
\usepackage{epstopdf}

\usepackage{wrapfig}

\newcommand{\R}{{\mathbb{R}}}

\usepackage[many]{tcolorbox}
\tcbuselibrary{skins}

\usepackage{tikz,circuitikz}
\usetikzlibrary{patterns,calc,angles,quotes,automata,positioning,decorations.markings,arrows,intersections,circuits.logic.US,shapes.arrows,shapes.gates.logic.US,backgrounds}
\usetikzlibrary{automata,positioning,decorations.markings,arrows,intersections,calc,shapes}

\colorlet{darkgreen}{green!40!black}
\colorlet{darkblue}{blue!60!black}
\colorlet{darkred}{red!50!black}
\colorlet{safecellcolor}{yellow!5}
\colorlet{goodcellcolor}{green!10}
\colorlet{badcellcolor}{blue!10}

\tikzset{
	>=stealth,
	box state/.style={draw,rounded corners,rectangle,minimum size=8mm},
	prob state/.style={draw,very thick,shape=circle,darkblue,minimum size=3mm,inner sep=0mm},
	node distance=2cm,on grid,auto, initial text=,
	every loop/.style={shorten >=0pt},
	accepting/.style={double distance=1.2pt, outer sep = 0.6pt+\pgflinewidth},
	accepting dot/.style={above=-2.5pt,circle,fill,darkgreen,inner sep=2pt,radius=1pt},
	loop above/.append style={every loop/.append style={out=120, in=60, looseness=5}},
	loop below/.append style={every loop/.append style={out=300, in=240, looseness=5}},
	loop left/.append style={every loop/.append style={out=210, in=150, looseness=5}},
	loop right/.append style={every loop/.append style={out=30, in=330, looseness=5}},
	accepting arc/.style={dashed},
	marked/.style={
		dashed,
		opacity=0.3
	},
	marked on/.style={alt=#1{marked}{}},
}
\ctikzset{bipoles/not port/width=0.75,bipoles/not port/height=0.5}
\ctikzset{bipoles/not port/circle width=0.3}
\ctikzset{tripoles/american nand port/circle width=0.2}

\usepackage{algorithm}
\usepackage{algorithmic}

\usepackage{xspace}

\def\BibTeX{{\rm B\kern-.05em{\sc i\kern-.025em b}\kern-.08em
		T\kern-.1667em\lower.7ex\hbox{E}\kern-.125emX}}

\usepackage{fancyhdr}

\newenvironment{nouppercase}{%
	\renewcommand{\uppercasenonmath}[1]{}}{}

\newcommand\scalemath[2]{\scalebox{#1}{\mbox{\ensuremath{\displaystyle #2}}}}

\usepackage{pdflscape}

\begin{document}

\begin{abstract}
This work is concerned with an assume-guarantee approach to compositionally control a New England 39-bus Test System (NETS). The proposed scheme is based on the new notion of \emph{robust simulation functions with disturbance refinement} alongside the composition of multiple subsystems to tackle the difficulties associated with scalability, also known as the \emph{curse of dimensionality}. In our proposed setting, we approximate concrete subsystems with abstractions that have lower
dimensions (\emph{a.k.a.} reduced-order models) while providing mathematical guarantees on controller synthesis through the quantification of an upper bound on the closeness between output trajectories of original systems and their abstractions. We propose two control methods to provide guarantees for NETS: one using the principle of interconnected synchronous machines and another considering the power flows in the network between neighbouring subsystems.
\end{abstract}

\title{{\LARGE{Formal Control of New England 39-Bus Test System: An Assume-Guarantee Approach}}}

\author{{\bf {\large Ben Wooding}}}
\author{{\bf {\large Abolfazl Lavaei}}}
\author{{\bf {\large Sadegh Soudjani}}\\
{\normalfont School of Computing, Newcastle University, United Kingdom}\\
\texttt{\{b.wooding1,abolfazl.lavaei,sadegh.soudjani\}@newcatle.ac.uk}}

\pagestyle{fancy}
\lhead{}
\rhead{}
\fancyhead[OL]{Ben Wooding, Abolfazl Lavaei, and Sadegh Soudjani}

\fancyhead[EL]{Formal Control of New England 39-Bus Test System: An Assume-Guarantee Approach}
\rhead{\thepage}
\cfoot{}

\begin{nouppercase}
	\maketitle
\end{nouppercase}

 \section{Introduction}
\label{sec:Introduction}

Cyber-physical systems (CPS) combine both cyber and physical components in interconnected models with interactions through feedback loops~\cite{lee2008cyber}. They are an important modelling framework for engineering real-life systems such as autonomous vehicles, medical devices and power systems, to name a few. The interconnection of these components in the models often results in high-dimensional systems with complex behaviour specifications that are generally safety critical in nature. Providing guarantees on the behaviour of these systems is therefore essential but also incredibly challenging. To tackle this difficulty, formal methods have been introduced in the relevant literature as a strong mathematical framework to provide guarantees on either verification or controller synthesis of CPS~\cite{Pnueli77,Lavaei_Survey}.

Symbolic control is one of the promising techniques for formal control synthesis of CPS~\cite{tabuada2009verification}. In particular, symbolic models (\emph{a.k.a.} finite abstractions) replace concrete systems to provide an easier medium for synthesis of a formal controller. In abstraction-based techniques, each discrete state and input in the finite abstraction maps to an aggregate collection of continuous states in the original (concrete) model. By establishing a similarity relation between original systems and their symbolic models, one can consider the abstract system as an appropriate substitute in the controller
design process with lower computational complexity while still
preserving closeness guarantees between the two systems.

Simulation and bisimulation functions are powerful techniques to relate output trajectories of abstract systems to
those of concrete ones~\cite{tabuada2009verification,baier2008principles}. If a concrete system is (bi)similar to an abstract system, only the abstract system needs to be considered in the formal synthesis process, while guarantees are still provided. For control systems where output trajectories of two systems may not be identical, \emph{approximate} (bi)simulation functions~\cite{girard2009hierarchical} have been developed in which output trajectories of two systems are only required to remain measurably close. In this case, the closeness between output trajectories can be bounded by some maximal $\epsilon$, known as the simulation relation error. Given an $\epsilon$-closeness, interface functions can be used to map the synthesized controller from the abstract system back to the concrete one. In~\cite{kurtz2020robust}, this type of relations is extended to \emph{robust} simulation functions (RSF) with small disturbances inside the concrete system, but with an unperturbed abstract system. 

Abstraction-based techniques often suffer severely from the \emph{curse of dimensionality} while dealing with high-dimensional systems~\cite{hsu2018multi}. To alleviate this computational complexity, one potential approach is to use compositional techniques: decompose a large-scale system into multiple subsystems and provide analysis over the high-dimensional system via its smaller subsystems~\cite{kerber2010}. Assume-guarantee contracts have been explored extensively in the literature to provide control techniques over a network of continuous-time dynamical systems~\cite{saoud2021assume}. Compositional approaches have also been used for the construction of (in)finite abstractions for interconnected systems based on abstractions of smaller subsystems~\cite{pola2016symbolic,tazaki2008bisimilar,lavaei2018CDCJ,lavaeiTAC2022,lavaei2020compositional,lavaei2019automated}.

Power networks are a demanding application of CPS that have received remarkable attention in the past decade. In particular, as the contribution of renewable energy rises, power networks are becoming increasingly intermittent. To ensure stability and functionality of  power networks, demand-side control techniques are required~\cite{bevrani2014robust}. In this respect, smart grid control involves the demand-side of a power grid responding to events in order to reduce the strain on the network, while also optimising consumer satisfaction and other specialist requirements~\cite{kundur1994}. Smart grids contain sensors and information-based technical devices, so it assumes that the current frequency, power generation or load values applied in different locations of the system can be accurately measured.

Formal methods play a significant role in power systems to provide formal analysis over this type of demanding systems. In this regard, the work~\cite{Stankovic2015approximate} proposes approximate bisimulations in transient power systems and employs differential-algebraic equations (DAEs) to model the New England 39-Bus Test System.
In~\cite{Althoff2014solo}, DAEs are utilised as models of the IEEE 14-Bus System and the IEEE 30-Bus System to provide reachability analysis for transient stability without performing any controller synthesis. The work~\cite{Li2019}, studies formal analysis of power systems via reachable sets of microgrids with distributed energy resources. The results of~\cite{Zonetti2019} use contract-based symbolic controller design for voltage stability in DC microgrids.

\noindent\textbf{Original Contributions.} In this work, we generalise the notion of robust simulation functions (RSFs) with disturbance refinement from linear systems to a class of nonlinear systems. We also provide an assume-guarantee contracts approach with RSF for the control of an interconnected network composed of several subsystems. Given the employed assume-guarantee contracts with RSF, we demonstrate the efficacy of our results on the New England $39$-bus Test System (NETS), as a large closely-coupled benchmark test system, composed of three $9$-dimensional subsystems (totally $27$ dimensions). We leverage model-order reduction techniques and construct a $3$-state reduced-order model for each subsystem (totally $9$ dimensions) to further mitigate the curse of dimensionality. We also provide a set of temporal logic specifications for the Great Britain power network using \emph{linear temporal logic} (LTL)~\cite{baier2008principles}.
We demonstrate our results for primary frequency control using two scenarios: (i) leveraging the principle of interconnected synchronous machines to control isolated subsystems, and (ii) considering internal disturbances in the network between different subsystems to provide accurate controls using shared information of neighbouring frequencies. For the sake of better illustrations of the results, we present our complex case study as a running example throughout the paper. We also use \emph{area} and \emph{subsystem} interchangeably throughout the paper.

A limited subset of our proposed results has recently been presented in~\cite{wooding2023ecc}. The results of this work differ from~\cite{wooding2023ecc} in three main directions. First and foremost, instead of considering only a single subsystem of NETS, we study an interconnected network of these subsystems that are then controlled compositionally. Secondly, we generalise the theoretical results of~\cite{wooding2023ecc} from simple linear systems to a class of nonlinear control systems. Finally, we apply our results to the New England 39-bus Test System, as a highly challenging large-scale closely-coupled system, which is significantly more complex than the case study in~\cite{wooding2023ecc}. The approach of~\cite{wooding2023ecc} cannot cope with the case study in this paper due to the scalability limitations caused by the curse of dimensionality. In addition, we
provide the proofs of all statements and the values used for simulation in the appendix, which were omitted in~\cite{wooding2023ecc}.

The paper is organised as follows. Preliminaries and the class of systems are provided in Section~\ref{sec:prel}, as well as introducing NETS as a running case study through the paper. We define the GB power network frequency specifications, expressed in LTL in Section~\ref{sec:PSspecs}. We generalise the notion of RSF with disturbance refinement to a class of nonlinear systems in Section~\ref{sec:simFun} and provide a proof of concept for the proposed technique in Section~\ref{sec:PoC}. We present the interconnection of subsystems in Section~\ref{sec:decomposition} and the methodology of assume-guarantee contracts in Section~\ref{sec:agc}. Demonstration of the proposed approaches for isolated areas and for compositional techniques with internal disturbances is provided in Sections~\ref{sec:isolatedControllers} and~\ref{sec:CompwII}, respectively. Finally, concluding remarks and future directions are provided in Section~\ref{sec:conclusion}.

\section{Notations and Preliminaries}
\label{sec:prel}

We employ the following notation throughout the paper. We denote the set of natural numbers, real, and non-negative real numbers with, $\mathbb{N}$, $\mathbb{R}$ and $\mathbb R^+$, respectively. The empty set is denoted by $\emptyset$. A function $\gamma:\reals^+\xrightarrow[]{}\reals^+$ is a class-$\kappa$ function if $\gamma$ is continuous, strictly increasing and $\gamma(0) = 0$. We use $\lvert\cdot\rvert$ for the absolute value, $\lVert a\rVert$ for the Euclidean norm of a vector $a$, and $\lVert a \rVert_\infty$ for taking the Euclidean norm followed by a maximisation over the bounded domain of $a$. $\mathbb{I}^n$ denotes an identity matrix in $\mathbb{R}^{n\times n}$, and $a \ll b$ represents $a$ much less than $b$. We denote intervals as subsets of real numbers by $\mathcal{B} =[\underline{\mathcal{B}},\overline{\mathcal{B}}]$ where $\underline{\mathcal{B}}$ and $\overline{\mathcal{B}}$ are used for the lower and upper boundaries of the interval. Specifically, $\mathcal{B}$ will denote the safe set and $\mathcal{T}$ the target set. 
All derivatives are taken with respect to time, and we often omit time for simplicity (\emph{e.g.,} use $\x$ instead of $\x(t)$). Given functions $g^i : X^i \rightarrow Y^i$, for any $i\in\{1,\ldots,N\}$, their Cartesian product $\Pi^N_{i=1}g^i:\Pi^N_{i=1}X^i\rightarrow\Pi^N_{i=1}Y^i$ is defined as $(\Pi^N_{i=1}g^i)(x^1,\ldots,x^n)=[g^1(x^1);\ldots;g^N(x^N)]$. We represent systems using $\Sigma$, where superscripts are used to label subsystems (\emph{i.e.,} $\Sigma^i$) and subscripts to represent systems and their abstractions; \emph{i.e.,} original system ($\Sigma_1$) and its reduced-order abstract system ($\Sigma_2$). 

\begin{definition}[Subsystems]\label{Def_Sub}
Consider a network of $N$ subsystems, where each subsystem $i$ can be modelled by $\Sigma^i_z = (X^i_z,U^i_z,V^i_z,W^i_z,g^i_z,Y^i_{z_1},Y^i_{z_2},h^i_{z_1},h^i_{z_2})$, $z\in\{1,2\}$, and $i\in\{1,\ldots,N\}$,
where:
\begin{itemize}
    \item $X^i_z\subseteq\reals^{n^i_z}$ are state sets of subsystems;
    \item $U^i_z\subseteq\reals^{p^i_z}$ are control input sets of subsystems;
    \item $V^i_z\subseteq\reals^{q^i_z}$ are external disturbance sets of subsystems;
    \item $W^i_z\subseteq\reals^{r^i_z}$ are internal disturbance sets of subsystems;
    \item $g^i_z: X^i_z\times U^i_z\times V^i_z\times W^i_z \to X^i_z$ are transition maps describing the evolution of subsystems;
    \item $Y^i_{z_1}\subseteq\reals^{m^i}$ are external output sets of subsystems;
    \item $Y^i_{z_2}\subseteq\reals^{m^i}$ are internal output sets of subsystems;
    \item $h^i_{z_1}: X^i_z \to Y^i_{z_1}$ are external output maps of subsystems;
    \item $h^i_{z_2}: X^i_z \to Y^i_{z_2}$ are internal output maps of subsystems.

\end{itemize}
The evolution of subsystems can be characterised by
\begin{equation}
\label{eq:both_disturbed_systems}
    \Sigma^i_z\!: \begin{cases}
\dot{\x}^i_z = g^i_z(\x^i_z,\u^i_z, \v^i_z, \w^i_z),\\
\y^i_{z_1} = h^i_{z_1}(\x^i_z),\\
\y^i_{z_2} = h^i_{z_2}(\x^i_z),
\end{cases} z\in\{1,2\},~i\in\{1,\ldots,N\}.
\end{equation}
where $\x^i_z\in X^i_z$, $\y^i_{z_1}\in Y^i_{z_1}$, $\y^i_{z_2}\in Y^i_{z_2}$, $\u_z^i\in U^i_z$, $\w^i_z\in W^i_z$, and $\v^i_z\in V^i_z$, where $\v^i_z$ are measurable and potentially large.
\end{definition}

Without loss of generality, we consider $\Sigma^i_1$ as our original (concrete) subsystem  and $\Sigma^i_2$ as it (possibly) lower-dimensional abstraction (with $n^i_2 \leq n^i_1$). In the following, we present the definition of interconnected systems in which subsystems $\Sigma^i_z$ are connected with each other via internal disturbances $\w^i_z$.

\begin{definition}[Interconnected Systems]
Consider a network of $N$ subsystems $\Sigma^i_z$, as defined in Definition~\ref{Def_Sub}, with a coupling matrix $\mathcal{M}$ among them.
The interconnection of $\Sigma^i_z$ for any $i\in\{1,\ldots,N\}$, is the interconnected control system $\Sigma_z = (X_z,U_z,V_z,g_z,Y_z,h_z)$, denoted by $\mathcal{I}(\Sigma^1_z,\ldots,\Sigma^N_z)$, such that
$X_z:= \Pi^N_{i=1}X^i_z$, $U_z:= \Pi^N_{i=1}U^i_z$, $V_z:= \Pi^N_{i=1}V^i_z$, $g_z:= \Pi^N_{i=1}g^i_z$, $Y_z:= \Pi^N_{i=1}Y^i_{z_1}$, and $h_z:= \Pi^N_{i=1}h^i_{z_1}$, with internal disturbances constrained by 
      \begin{equation*}
          [\w^1_z;\ldots;\w^N_z] = \mathcal{M}[{\y}^1_{z_2};\ldots;{\y}^N_{z_2}].
      \end{equation*}
\end{definition}

The evolution of the interconnected system is therefore characterised by
\begin{equation*}
\label{eq:interconnected_disturbed_systems}
    \Sigma_z\!: \begin{cases}
\dot{\x}_z = g_z(\x_z,\u_z, \v_z),\\
\y_{z} = h_{z}(\x_z)
\end{cases} z\in\{1,2\}.
\end{equation*}

\smallskip
\noindent{\bf Linear Temporal Logic Specifications.}
For dynamical systems in~\eqref{eq:both_disturbed_systems}, we consider linear temporal logic (LTL) specifications with syntax \cite{baier2008principles}
\begin{equation*}
\psi :=  \operatorname{true} \,|\, p \,|\, \neg \psi \,|\,\psi_1 \wedge \psi_2 \,|\, \mathord{\bigcirc} \psi \,|\, \psi_1\mathbin{\sf U} \psi_2,
\end{equation*}
where $p$ is the element of an atomic proposition. Let $\word$ be an infinite word, that is, a
string composed of letters from power sets of the atomic proposition, and $\word_k$ be a subsequence
(suffix) of $\word$. Then  
the satisfaction relation between $\word$ and a property $\psi$, expressed as LTL,  is denoted by $\word\vDash\psi$. Furthermore,
$\word_k\vDash \neg \psi$  if $\word_k\nvDash\psi$ and 
we say that  $\word_k\vDash \psi_1\wedge\psi_2$ 
if $ \word_k\vDash \psi_1$ and $\word_k\vDash \psi_2$.
The next operator $\word_k\vDash\mathord{\bigcirc}\psi $ holds if the property holds at the next time instance.
The temporal until operator $\word_k\vDash \psi_1\mathbin{\sf U}\psi_2$  holds if $ \exists i \in \mathbb{N}:$ $\word_{k+i} \vDash \psi_2, \mbox{and } 
\forall j \in{\mathbb{N}:} 0\leq j<i, \word_{k+j}\vDash \psi_1
$.
Disjunction ($\vee$) can be defined by
$ \word_k\vDash \psi_1\vee\psi_2\ \Leftrightarrow  \  \word_k\vDash \neg(\neg\psi_1 \wedge \neg\psi_2)$. The operator $\word_k\vDash\LTLeventually \psi$ is used to denote that the property will eventually happen at some point in the future. The operator $\word_k\vDash\LTLalways \psi$ signifies that $\psi$ must always be true at all times in the future. Additionally, we use a subscript under an operator to indicate the time horizon over a specification, \emph{e.g.,} $\word_k\vDash\LTLeventually^{30}\psi$ would signify the property will eventually hold before $30$ units of time. 

\noindent\textbf{Running Case Study.} We apply our developed approach in this work mainly to a model of New England $39$-bus Test System (NETS) as a highly challenging and demanding system. This model is similar to the three-control area power system in \cite{oshnoei2021novel}, where we present it as a running case study throughout this work for the sake of better illustration. NETS has $10$ machines, $39$ buses, $46$ lines and three areas. One of the generators is used to represent the connection between the NETS power system and the wider American power network.  This provides an especially challenging case study for the techniques we propose in this work.

NETS can be decomposed into three smaller areas (\emph{a.k.a.} subsystems), each of which contains three generators. The interconnected system consists of $27$ states, with $9$ states in each area. In addition, we consider possible external disturbances together with inputs (one per area)
that can be used for control synthesis purposes, \emph{e.g.,} Energy Storage Systems (ESSs) or Plug-in Electric Vehicles (EVs). We define internal disturbances as power dynamics affecting a local area, \emph{i.e.,} own subsystem, caused by neighbouring areas \emph{i.e.,} other subsystems.

In this running case study, the main goal is to formally control NETS. Given that formal control approaches often struggle with scalability, we employ model order reduction techniques  together with RSF with disturbance refinement to reduce the number of states while providing mathematical guarantees for the system behaviour. Reducing NETS from $27$ states to lower dimensions introduces a reduction error $\epsilon$ which is generally very large. Hence, we employ a compositional technique to first decompose the NETS into three $9$-dimensional areas and then reduce the dimension of each area via constructing reduced-order abstractions. 

It is worth highlighting that although each decomposed area of NETS has $9$ states, this is still an intractable problem due to occurring curse of dimensionality during synthesis procedure. To resolve this issue, we first aim at building a reduced-model abstraction with $3$ states for each area and then constructing an RSF with disturbance refinement as a relation between each concrete area and its reduced-order model. To maintain the interconnection of all areas, the frequency of neighbouring areas is used as internal disturbances of the local area. We use the following relationship between frequencies to connect these areas

\begin{equation}
\label{eq:freq-interface}
    \frac{2\pi}{s} \sum^{N}_{j=1} T^{ij} (f^i- f^j),
\end{equation}
where $f^i$ is the local area frequency
and $f^j$ are neighbouring area frequencies, $T^{ij}$ are constants related to the power interchange between the respective neighbours. By leveraging the principle of interconnected synchronous machines~\cite{kundur1994}, one can assume that for all neigbours, $f^i = f^j$. This assumption results in~\eqref{eq:freq-interface} to be zero so that each area is simplified to $9$ states with no internal disturbances. The linear dynamics of NETS are acquired using the Simulink Model Linearizer on the closed-loop system.

\section{Frequency Specifications}
\label{sec:PSspecs}
In this section, we define the requirements on frequency regulation of the GB power grid using LTL specifications. The requirements are compiled from the collection of references~\cite{greenwood2017frequency,TheGridCode2020,NETS2019,Smith2016,FirmFrequencyResponse17,EnhancedFrequencyResponse} as follows. The nominal frequency of the GB power grid is $f_0=50~Hz$. The frequency $f$ should remain between the statutory limits $\mathcal{S} = [\underline{\mathcal{S}},\overline{\mathcal{S}}]$ with $\underline{\mathcal{S}} = 49.5~Hz$ and $\overline{\mathcal{S}} = 50.5~Hz$, for all losses up to the maximum \emph{normal infeed loss} ($\mathcal{L}= 1320~MW$):
\begin{equation*}
\label{eq:spec_normalloss}
    \psi_{normal} :=~ [loss \leq \mathcal{L}] \implies \LTLalways [f\in\mathcal{S}].
\end{equation*}
Losses greater than $1320~MW$ are considered \emph{infrequent infeed losses}  and may fall below the statutory limits briefly, but no lower than the containment zone value $\mathcal{Z}=49.2~Hz$. Within a time constraint of $60$~seconds, the frequency should return to the statutory limits under the following conditions:
\begin{align*}
\label{eq:spec_infrequentloss}
    \psi_{infrequent} :=~&[loss \geq \mathcal{L} \wedge f<\underline{\mathcal{S}}] \implies
    [\LTLeventually^{60} (f\in\mathcal{S})\wedge\LTLalways (f\geq\mathcal{Z})].
\end{align*}
If the frequency rises above $52~Hz$ or falls below $47~Hz$, \emph{i.e.,} there is system shutdown, which should be avoided at all costs:
\begin{equation*}
\label{eq:spec_shutdown}
    \psi_{shutdown} :=~ \LTLalways[47 \leq f\leq 52].
\end{equation*}
Additionally, the GB power grid specifies certain minimum time constraints for the devices contributing to primary, secondary and high frequency response services, as discussed next.

\noindent\textbf{Firm Frequency Response (FFR).} For devices contributing to \emph{primary frequency response}, it is necessary to inject power ($\mathcal{P}_p > 0~MW$) within 2 seconds of a \emph{low frequency event} ($\mathcal{E}_{low}$), and provide maximum power ($\mathcal{P}_p^{max}$) by 10 seconds. This maximum power must be at least $1~MW$ per response device or aggregated load. This delivery should be maintained for 30 seconds:
\begin{align}
    \label{eq:primary_frequency_response}
    \psi_{p1} :=~ &\mathcal{E}_{low} \implies
    [\LTLeventually^2(\mathcal{P}_p > 0)
    \wedge\LTLeventually^{10}(\mathcal{P}_p=\mathcal{P}_p^{max})], \nonumber\\
    \psi_{p2} :=~ &[\mathcal{P}_p = \mathcal{P}_p^{max}] \implies
    \LTLalways^{30}[\mathcal{P}_p = \mathcal{P}_p^{max}],\nonumber\\
    \psi_{p} :=~ &\psi_{p1} \wedge \psi_{p2}.
\end{align}
For devices contributing to \emph{secondary frequency response}, it is essential to begin injecting maximum power ($\mathcal{P}_s^{max}$) within $30$ seconds of a low frequency event. Similarly, the delivery ($\mathcal{P}_s$) should be maintained for 30 minutes:
\begin{align}
    \label{eq:secondary_frequency_response}
    \psi_{s1} :=~ &\mathcal{E}_{low} \implies
    \LTLeventually^{30}[\mathcal{P}_s=\mathcal{P}_s^{max}], \nonumber\\
    \psi_{s2} :=~ &[\mathcal{P}_s = \mathcal{P}_s^{max}] \implies
    \LTLalways^{1800}[\mathcal{P}_s = \mathcal{P}_s^{max}],\nonumber\\
    \psi_{s} :=~ &\psi_{s1} \wedge \psi_{s2}.
\end{align}
It is possible for devices to perform both primary and secondary response services in the power grid with the following specification:
\begin{equation*} \label{eq:joint_primary_secondary_frequency_response}
    \psi_{ps}:=~\psi_p\wedge\psi_s.
\end{equation*}
Equivalent \emph{high frequency response} specifications for any \emph{high frequency event} ($\mathcal{E}_{high}$) are similar to both~\eqref{eq:primary_frequency_response} and~\eqref{eq:secondary_frequency_response} but without a fixed delivery duration. 

\noindent\textbf{Enhanced Frequency Response (EFR).}
Taking advantage of the fast response capabilities of\emph{ energy storage systems} (ESSs), enhanced frequency response (EFR) is designed to allow \emph{state-of-charge} (SoC) management which is not possible with FFR. ESS should respond within $1$ second of the frequency crossing the deadband threshold which can be set at $db=[49.95,50.05]~Hz$ for a wide deadband and $db=[49.985,50.015]~Hz$ for a narrow deadband. The EFR service must be bidirectional, \emph{i.e,} both exported and imported to/from the grid. It must be possible for the EFR service to be provided at $100\%$ capacity ($\mathcal{P}_{EFR}^{max}$) for a minimum of 15 minutes. To avoid short-term frequency instability issues
from the fast response, ramp-rate limitations 
have been included in the specification when the frequency is inside the envelope but outside of the deadband. The ramp-rate limitations are included to limit short-term stability problems~\cite{greenwood2017frequency}.
The maximum change in output is limited as a proportion of the rate of change of the frequency (RoCoF or $\frac{\partial f}{\partial t}$). The ramping constant $k$ is $0.45$ for the wide deadband and $0.485$ for the narrow deadband:
\begin{align*}
    &\mathcal{P}_{EFR}^{max}(-\frac{1}{k}\frac{\partial f}{\partial t}-0.01) < \frac{\partial \mathcal{P}}{\partial t} < \mathcal{P}_{EFR}^{max}(-\frac{1}{k}\frac{\partial f}{\partial t}+0.01),\nonumber\\
    &\psi^{1}_{efr} :=~[f\notin db] \implies
    \LTLeventually^1[\ \mathcal{P}=\mathcal{P}_{EFR}^{max}], \nonumber\\
    &\psi^{2}_{efr} :=~[\mathcal{P}=\mathcal{P}_{EFR}^{max}] \implies
\LTLalways_{900}[\mathcal{P}=\mathcal{P}_{EFR}^{max}], \nonumber\\
    &\psi_{efr}:=\psi^{1}_{efr}\wedge\psi^{2}_{efr}.
\end{align*}

\noindent\textbf{{Running case study (continued).}}
We consider a stricter primary frequency specification, in which the frequency $f$ can deviate away from its steady state value $f_0$, the deviation is denoted by $\Delta f = f-f_0$.
We bound two regions that the frequency deviation should never transition into, $\mathcal{A}_{ub} = (\overline{\mathcal{B}},+\infty)$ and $\mathcal{A}_{lb} = (-\infty,\underline{\mathcal{B}})$.
Additionally, whenever there are deviations, the frequency should return to the target range $\mathcal{T} = [\underline{\mathcal{T}}, \overline{\mathcal{T}}]$.
The desired system behaviour can be described by the following LTL formulae:
\begin{align*}
   &\psi = \LTLalways(\psi_1 \land \psi_2) \text{ with }\psi_1 = \LTLeventually\mathcal{T},~\psi_2 = \lnot( \mathcal{A}_{ub} \lor \mathcal{A}_{lb}).
\end{align*}
We modify this specification appropriately with the simulation relation error $\epsilon$ from~\eqref{eq:SRE}, to acquire a conservative specification $\hat{\psi}$ over $\Sigma_2$ as:
\begin{equation}
     \hat{\psi} = \LTLalways(\hat{\psi}_1 \land \hat{\psi}_2) \text{ with }\hat{\psi}_1 = \LTLeventually\hat{\mathcal{T}},
    ~\hat{\psi}_2 = \lnot( \hat{\mathcal{A}}_{ub} \lor \hat{\mathcal{A}}_{lb}),
    \label{eq:abs_spec}
\end{equation}
with $\hat{\mathcal{T}} = [\underline{\mathcal{T}}+\epsilon, \overline{\mathcal{T}}-\epsilon]$, $\hat{\mathcal{A}}_{ub} = (\overline{\mathcal{B}}-\epsilon,+\infty)$ and $\hat{\mathcal{A}}_{lb} =( -\infty,\underline{\mathcal{B}}+\epsilon)$. This modification ensures that whenever the abstract system $\Sigma_2$ satisfies $\hat{\psi}$, the concrete system $\Sigma_1$ satisfies the original specification $\psi$ by applying appropriate input and disturbance interface functions for refining the controller.

\section{Simulation Functions}
\label{sec:simFun}
In this work, we leverage the notion of robust simulation functions to construct an abstract system which is $\epsilon$-close to the concrete one, where $\epsilon$ remains small enough. In the following subsection, we show how incorporating the disturbance of the concrete system into the abstract one, through an interface function $d_\simF$, can further reduce the simulation relation error $\epsilon$ between $\Sigma_1$ and $\Sigma_2$. This enables one to perform controller synthesis on the abstract domain and refine it back over potentially high-dimensional original system while improving the scalability of the control scheme.

\subsection{Robust Simulation Function with Disturbance Refinement}
Given the system in~\eqref{eq:both_disturbed_systems}, we formalize the definition of a robust simulation function $\simF$ with two interface functions $u_\simF$ and $d_\simF$ as the following.
\smallskip
\begin{definition}[Robust Simulation Functions]
\label{def:Robust_Simulation_Function_Final}
Consider two systems of the form \eqref{eq:both_disturbed_systems}. Let $\simF: X_1\times X_2\xrightarrow[]{}\reals^+$
be a smooth function, $u_\simF:U_2\times X_1\times X_2\xrightarrow[]{}U_1$ and $d_\simF: V_1\times W_1\times X_1\times X_2\xrightarrow[]{}V_2\times W_2$ be continuous functions. Then the function $\simF$ is called a robust simulation function (RSF) from $\Sigma_2$ to $\Sigma_1$
and $u_\simF$, $d_\simF$ are its associated interface functions if there exist class-$\kappa$ functions $\gamma_1$ and $\gamma_2$ such that for all $\x_1\in X_1$, $\x_2\in X_2$
,
\begin{equation}
\label{eq:error-bound}
    \lVert h_1(\x_1) - h_2(\x_2) \rVert \leq \simFfull,
\end{equation}
for any $\u_2\in U_2$ and $\d_1\in [V_1~W_1]^T$ satisfying $\gamma_1(\lVert \d_1 \rVert) + \gamma_2(\lVert \u_2 \rVert) \leq \simFfull$, we have
\begin{align}
\label{eq:decay-rate}
    &\frac{\partial\simF}{\partial \x_2}f_2(\x_2,\u_2,d_\simF(\d_1,\x_1,\x_2)) +
    \quad\frac{\partial\simF}{\partial \x_1}f_1(\x_1,u_\simF(\u_2,\x_1,\x_2),\mathbf{d}_1) \leq 0.
\end{align}
We say $\Sigma_1$ robustly approximately simulates $\Sigma_2$ if there exists an RSF $\simF$ from $\Sigma_2$ to $\Sigma_1$.
\end{definition}
\smallskip

\begin{comment}
\begin{remark}
Definition~\ref{def:Robust_Simulation_Function_Final} is a generalisation of the robust approximate simulation function proposed in~\cite{kurtz2020robust}. In particular, when $d_\simF = 0$, then the existing robust approximate simulation in~\cite{kurtz2020robust} is recovered.
\end{remark}
\end{comment}
In the next subsection, we focus on a class of nonlinear control systems with potentially large measurable disturbances and propose an approach to construct its reduced-dimensional abstractions together with an RSF as presented in Definition~\ref{def:Robust_Simulation_Function_Final}.

\subsection{Nonlinear Systems under Large Measurable Disturbance}
Here, we focus on a class of nonlinear control systems with (potentially large) measurable disturbances. A model in this class and its abstraction are specified by
\begin{equation}
\label{eq:nonlinear_both_disturbed_systems}
\Sigma_z : \begin{cases}
\dot{\x}_z = &A_z\x_z + B_z\u_z + G_{z}\v_z + S_{z}\w_z + E_z\phi(F_z\x_z),\\
\y_z = &C_z\x_z,
\end{cases}
\end{equation}
\noindent where
$z\in\{1,2\}$ represents the model and its abstraction respectively, and where $A_z \in \reals^{n_z\times n_z}, B_z \in \reals^{n_z\times p} ,C_z \in \reals^{m\times n_z}, G_{z} \in \reals^{n_z\times q}, S_{z} \in \reals^{n_z\times r}, E_z \in \reals^{n_z\times 1}, F_z \in \reals^{1\times n_z}$. In addition, $\phi:\reals \rightarrow \reals$ is a nonlinear term satisfying the following slop restriction:
\begin{equation}
\label{eq:slope_restriction}
    a \leq \frac{\phi(c)-\phi(d)}{c-d} \leq b,\quad \forall c,d\in\reals,c\neq d.
\end{equation}

\begin{remark}
 Note that if $E_1 = 0$ and/or $F_1 = 0$ in~\eqref{eq:nonlinear_both_disturbed_systems}, our proposed approach simplifies to the one we provided in~\cite{wooding2023ecc} for the class of linear control systems with potentially large measurable disturbances.
\end{remark}

We define $\d_z=[\v_z~\w_z]^T\in[V_z~W_z]^T$ and $D_z = [G_z~S_z]\in\reals^{n_z\times(q+r)}$ as the concatenation of external and internal disturbances. It is assumed $\d_1$ is a measured disturbance having some known bound $\lVert\d_1\rVert_{\infty} \leq d_{\max}$. Moreover, $\d_2$ is derived from $\d_1$ with the interface function $d_\simF$ (cf. \eqref{eq:interface_d}). We now present the main
problem that we aim to solve in this work.
\begin{resp}
\begin{problem}\label{Prob1}
Given a nonlinear system $\Sigma_1$ under large measurable disturbances and an LTL specification $\psi$, construct its reduced-dimensional abstraction $\Sigma_2$ together with an RSF as presented in Definition~\ref{def:Robust_Simulation_Function_Final}. Leverage the constructed abstraction $\Sigma_2$ and design a formal controller through simulation relations with disturbance refinement such that the specification is satisfied over the original system. We assume that the (potentially) large disturbance $\v_1$ is measurable with a known upper bound.
\end{problem}
\end{resp}
In order to address Problem~\ref{Prob1}, we raise the following lemma and theorems.

\begin{lemma}
\label{lem:M_and_K}
If $\Sigma_1$ is stabilisable, there are matrices $K_1,K_2,P,D_2,Q_1,L_{11},L_{21}$ such that $H$ is Hurwitz, and there exist a positive-definite matrix $M$ and a positive constant $\lambda$ such that the following matrix inequalities hold:
\begin{subequations}
\label{eq:LMIs}
\begin{align}
\label{eq:Mhold1}
    &C_1^TC_1 \leq M, \\
\label{eq:Mhold2}
    &H^TM + MH \leq -2\lambda M,
\end{align}
\end{subequations}
where 
\begin{align*}
&H = (A_1-PD_2(K_1+Q_1)+B_1K_2) +\bar{\delta}(E_1+B_1L_{21}-PD_2L_{11})F_1.
\end{align*}
Here $\bar\delta$ is the upper bound of $\delta$, where $\delta$
is a scalar in the interval $[a,b]$, in $\phi(F_1\x_1)-\phi(F_1P\x_2) = \delta F_1(\x_1-P\x_2)$ obtained from the slope restriction~\eqref{eq:slope_restriction}.
\end{lemma}

\smallskip
 Using Lemma~\ref{lem:M_and_K}, we now propose the next theorem to construct an RSF.
\begin{theorem}
\label{thm:one_final}
Consider two systems of the form \eqref{eq:nonlinear_both_disturbed_systems}. Assume that $\Sigma_1$ is stabilisable, a feedback gain $K_1$ exists for $\Sigma_2$ and that there exist matrices $P$, $K_2$, $Q_1$, $Q_2$, $L_{11}$, $L_{12}$, $L_{21}$ and $L_{22}$ such that the following matrix equalities hold:
\begin{subequations}
\label{eq:SDPs}
\begin{align}
\label{eq:C2=C1P}
    &C_2=C_1P,\\
\label{eq:F1=F2P}
&F_2 = F_1P, \\
\label{eq:newPA_2}
   &A_1P + B_1Q_2 = PA_2 + PD_2Q_1P, \\
\label{eq:E1=PE2}
&E_1 = PE_2 -B_1(L_{21}-L_{22})+PD_2(L_{11}-L_{12}).
\end{align}
\end{subequations}
Then $\mathcal V$ in the form of 
\begin{equation*}
    \simFfull = \sqrt{(\x_1-P\x_2)^TM(\x_1-P\x_2)}
\end{equation*}
is an RSF from $\Sigma_2$ to $\Sigma_1$ with its associated interfaces
\begin{subequations}
\label{eq:interfaces}
\begin{align}
\label{eq:interface_u}
    u_\simF &= K_2(\x_1-P\x_2) + Q_2\x_2+R_2\u_2 +L_{21}\phi(F_1\x_1)-L_{22}\phi(F_1P\x_2),\\
    \label{eq:interface_d}
    d_\simF &= K_1(\x_1-P\x_2) + Q_1\x_1+R_1\d_1+L_{11}\phi(F_1\x_1)-L_{12}\phi(F_1P\x_2).
\end{align}
\end{subequations}

In addition, the class-$\kappa$ functions $\gamma_1$ and $\gamma_2$ are designed as
\begin{align}\label{eq:gamma1_final}
\gamma_1(\nu) &= \frac{\lVert\sqrt{M}(D_1-PD_2R_1)\rVert}{\lambda}\nu, \\\label{eq:gamma2_final}
\gamma_2(\nu) &= \frac{\lVert\sqrt{M}(B_1R_2-PB_2)\rVert}{\lambda}\nu,
\end{align}
where $R_1$ and $R_2$ are some arbitrary matrices of appropriate dimensions, and $M, \lambda$ are matrices satisfying~\eqref{eq:LMIs}.
\end{theorem}
\smallskip

We now leverage the constructed $\mathcal V$ in Theorem~\ref{thm:one_final} and  quantify the mismatch between output trajectories of $\Sigma_1$ and $\Sigma_2$ with measurable disturbances, as presented in the next theorem.
\begin{theorem}
\label{thm:two_final}
Consider two systems of the form \eqref{eq:nonlinear_both_disturbed_systems}. Let $\simF$ be an RSF from $\Sigma_2$ to $\Sigma_1$ with its associated interface function $u_\simF$. Let $\u_2(t)$ be an admissible input of $\Sigma_2$ and $\x_1(t)$ be a state trajectory of $\Sigma_1$ satisfying
\begin{equation}
    \label{eq:thm2_eq}
    \dot{\x}_1 = A_1\x_1 + B_1u_\simF + D_1\d_1 + E_1\phi(F_1\x_1).
\end{equation}
Then
\begin{align}
    \lVert \y_1(t) - \y_2(t)\rVert \leq \epsilon = \max\{\simF(\x_1(0),\x_2(0)),\gamma_1(\lVert\d_1\rVert_\infty) + \gamma_2(\lVert\u_2\rVert_\infty)\}.
    \label{eq:SRE}
\end{align}
\end{theorem}

Our primary goal of employing RSF is to construct an abstract system $\Sigma_2$ which is $\epsilon$-close to the concrete system $\Sigma_1$, where $\epsilon$ remains small enough.
Note that in the modified specification \eqref{eq:abs_spec}, any value $\epsilon$ such that
$\hat{\mathcal{T}} = \emptyset$ causes the set of controllers enforcing the specification to also be empty. Therefore, our approximation approach must provide error thresholds small enough to give a feasible controller on the abstract system.

\section{Proof of Concept}
\label{sec:PoC}

We consider $\Sigma_1$ as the interconnected NETS, $\Sigma^i_1$ as the $9$-state decomposed subsystem $i$ and $\Sigma^i_2$ as the $3$-state reduced-order subsystem $i$. In all scenarios of this case study, we consider a power loss disturbance of $\v^i_1=1$ per unit ($100$~MW, equivalent to a typical generator or 35,000 households) as the default external disturbance. We construct abstract systems $\Sigma^i_2$ using MATLAB's \emph{balreal} function by truncating matrices to a reduced-state order of $3$. We employ YALMIP \cite{YALMIP} and MOSEK \cite{mosek} for solving LMIs on a macOS machine with $8$ GB RAM and Intel Core i$5$ Processor. We also use the tool SCOTS~\cite{Rungger2016} for the synthesis of the symbolic controller using a high-performance computer with $2$ nodes and $11$ GB memory per core. Simulations are run over a time horizon of $6$ seconds, with a time step of $0.005$ seconds. The values of the interconnected NETS model as well as the different subsystems can be found in the appendix.

\begin{figure}[h]
    \centering
    \vspace{0.5cm}
    \includegraphics[width=0.6\textwidth]{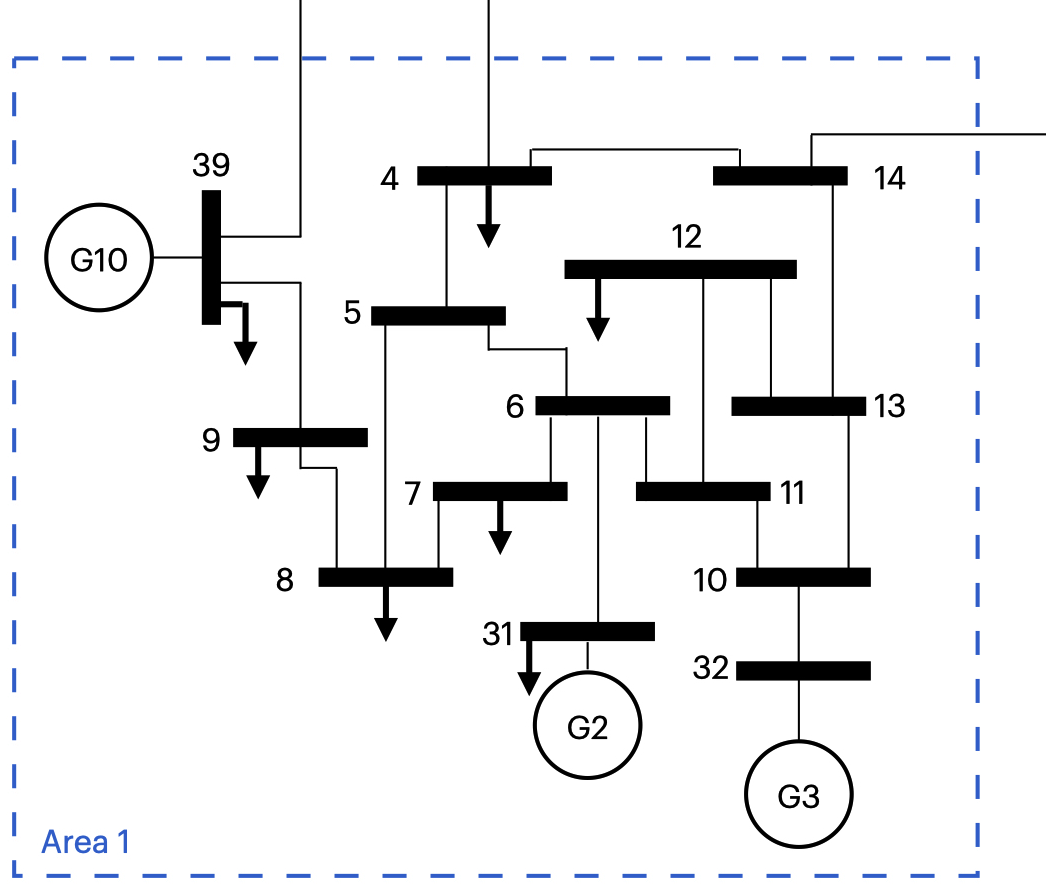}
    \caption{A single-line diagram of Area $1$ of New England $39$ Bus Test System, with generators (G$2$, G$3$, G$10$), buses (thick bars), loads (thick arrows), and power lines.}
    \vspace{0.5cm}
    \label{fig:39busSystem}
\end{figure}

\noindent\textbf{{Running case study (continued).}}
 To demonstrate the proposed RSF with disturbance refinement, we consider just one area of NETS, containing $9$ states with one input, one external disturbance and no internal disturbance. The single-line diagram for this system is depicted in Figure~\ref{fig:39busSystem}. A linear model for Area $1$ of NETS is acquired using the Simulink Model Linearizer on the closed-loop system.

To add nonlinear parts to this model, we consider a collection of energy storage systems (ESSs) which provide feedback control to the system depending on the current frequency. The power output of these ESSs is limited by a saturation function. We assume to have knowledge of dynamics of this feedback but no control over its power output. The dynamics of this output are adapted and simplified from aggregate battery charger models appeared in~\cite{Izadkhast2015,MUCHAOXU2018}:

\begin{equation*}
    \Delta P_{ESS} = N_{ESS} \times \mathsf{sat}(\frac{k_{ESS}}{R_{ESS}}\Delta f),
\end{equation*}
where $P_{ESS}$ is the power contribution of EVs, $f$ is the system frequency, $N_{ESS}$ is the number of participating ESSs, $k_{ESS}$ is the average participation factor, and $R_{ESS}$ is the droop constant. The saturation function $\mathsf{sat}(x)$ is also defined as

\begin{equation*}
    \mathsf{sat}(x) := \begin{cases}
    ESS_{max},\quad &x \geq ESS_{max}, \\
    x, \quad &ESS_{min} \le x \le ESS_{max}, \\
    ESS_{min}, \quad &x \leq ESS_{min}.
    \end{cases}
\end{equation*}
We assume that $\v_1$ is measurable given that the disturbance may represent changes in the behaviour of generation and load components, e.g., generators, plug-in electric vehicles (EVs) and ESSs. The generation or load values of these components may be known to operators and the connection/disconnection of these components could be tracked through sensors in a smart grid. We assume we have access to a fleet of EVs which can connect/disconnect from the power grid almost instantaneously. Such responsive loads are flexible and can be used for load shedding \cite{wooding2022control} and frequency regulation of smart grids \cite{wooding2020formal}. The dynamics of the model are therefore a nonlinear system $\Sigma^1_1$ equivalent to~\eqref{eq:nonlinear_both_disturbed_systems}.

\subsection{Simulation Relation Error}
\label{sub:uncontrolled}
\smallskip
\noindent\emph{Uncontrolled system.} If the response of EVs is not included in the system ($\u_1=0$), the open-loop nonlinear system $\Sigma_1$ has the maximum frequency deviation of $\Delta f = -0.6872 Hz$, which clearly violates the specification $\psi$. Therefore, the contribution of EVs is essential to satisfy the specification on the frequency.

\smallskip
\noindent\emph{Abstraction with disturbance refinement.}
We now use our proposed approaches from Theorems~\ref{thm:one_final}--\ref{thm:two_final} with the proposed disturbance interface function. The safe and target sets are defined as $\underline{\mathcal{B}} = -0.35$, $\overline{\mathcal{B}} = 0.5$, $\underline{\mathcal{T}} = -0.3$ and $\overline{\mathcal{T}} = 0.5$. We assume, $\lambda = 1.7$,
$\lVert\u_2\rVert_\infty = 0.5$,
and
$0.01\,\mathbb{I}_9 \leq \Mbar \leq120\,\mathbb{I}_9$, $L_{22} = 1$, $L_{21} = 0$, $D_2 = E_2$ and $Q_1=K_1=0$. We optimise $R_1$, $R_2$ and $B_2$ to minimise~\eqref{eq:gamma1_final} and~\eqref{eq:gamma2_final}, respectively. Accordingly, we get the value $\epsilon = 0.1019$.

\subsection{Controller Synthesis}
\begin{figure}[h]
\vspace{0.5cm}
    \includegraphics[width=0.6\textwidth]{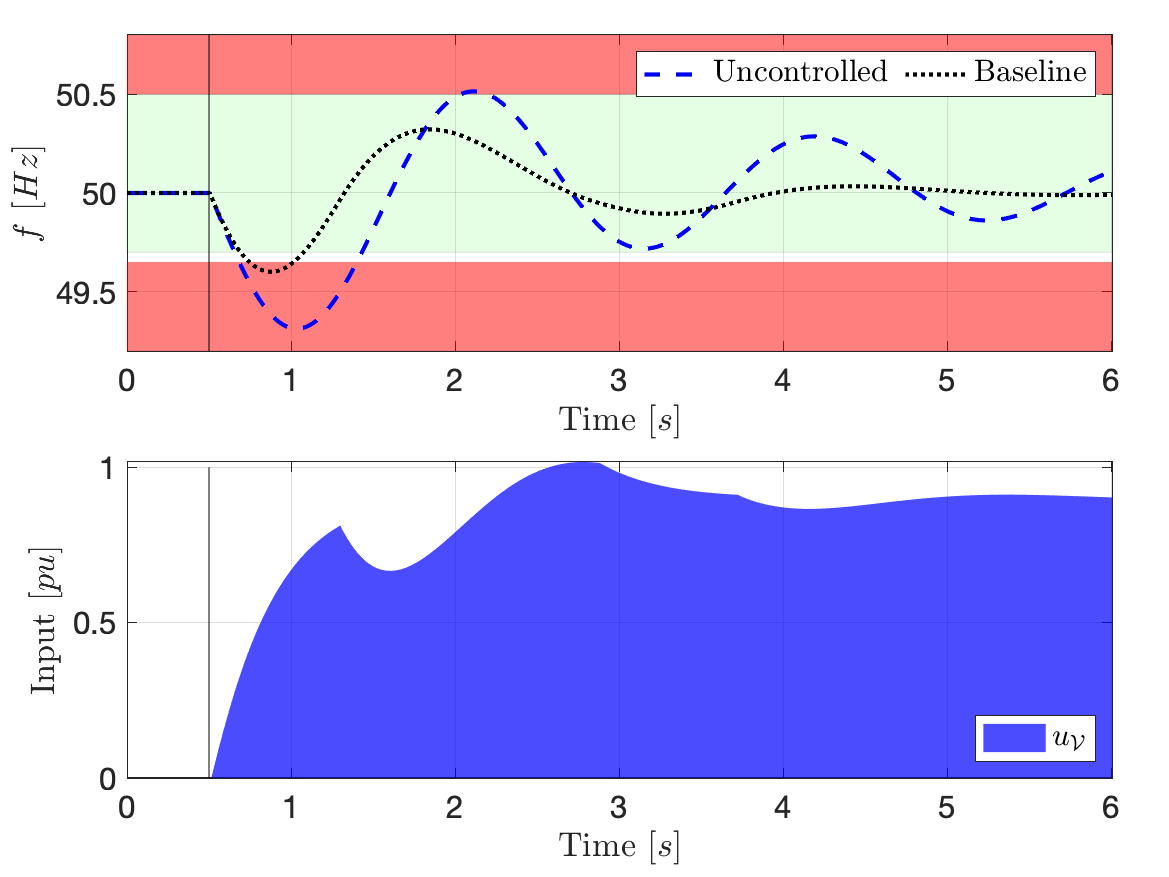}
    \centering
    \caption{\textbf{Top:} Target range $\mathcal{T}$ is shown in green, $\mathcal{A}_{ub}$ and $\mathcal{A}_{lb}$ are shown in red as two regions that the system should never transition into (unsafe regions). The baseline controller notably improves the frequency response of the system in compare with the uncontrolled system. However, both curves still fall into the red unsafe region. \textbf{Bottom:} The input $u_\simF$ is a byproduct of the simulation relation interface keeping $\Sigma_1$ and $\Sigma_2$ $\epsilon$-close. Since no controller is synthesised over $\Sigma_2$, then $\u_2 = 0$.}
    \vspace{0.5cm}
    \label{fig:open_loop_vs_closed_loop}
\end{figure}

\noindent\emph{Baseline controller.}
We consider our RSF with the constructed abstract system $\Sigma_2$ and the interface functions \eqref{eq:interfaces} but with $\u_2 = 0$ in \eqref{eq:interface_u}.
As $Q_2$ and $K_2$ are non-zero in~\eqref{eq:interface_u}, control inputs are chosen automatically based on the current states of $\Sigma_1$ and $\Sigma_2$ to maintain outputs of the two systems within distance $\epsilon$. When the power system frequency moves away from its steady-state value, the input interface function $u_\simF$ generates a control input for $\Sigma_1$, which is considered here as a \textit{baseline controller}.
The frequency response in $\Sigma_1$ without EV participation (uncontrolled system with $\u_1=0$) against the baseline controller is depicted in Figure~\ref{fig:open_loop_vs_closed_loop}. Although the baseline controller reduces the frequency deviations, it is still unable to satisfy the required specification $\psi$.

\noindent\emph{Controller using RSF.}
We employ the constructed abstraction $\Sigma_2$ as an appropriate substitute in the controller synthesis process. In particular, by knowing $\epsilon$ as the maximum error between outputs of $\Sigma_1$ and $\Sigma_2$, a symbolic controller can be first designed for the reduced-order model $\Sigma_2$ to satisfy $\hat\psi$ and then be refined back to $\Sigma_1$ while providing a guarantee on the satisfaction of $\psi$. The synthesis of the symbolic controller takes $77$ minutes and $34$ seconds.
\begin{remark}
Note that synthesising such a symbolic controller directly from any $9$-dimensional system is impossible due the required exponentially large computational time and memory space.
\end{remark}

\begin{figure}[h]
\vspace{0.5cm}
    \includegraphics[width=0.6\textwidth]{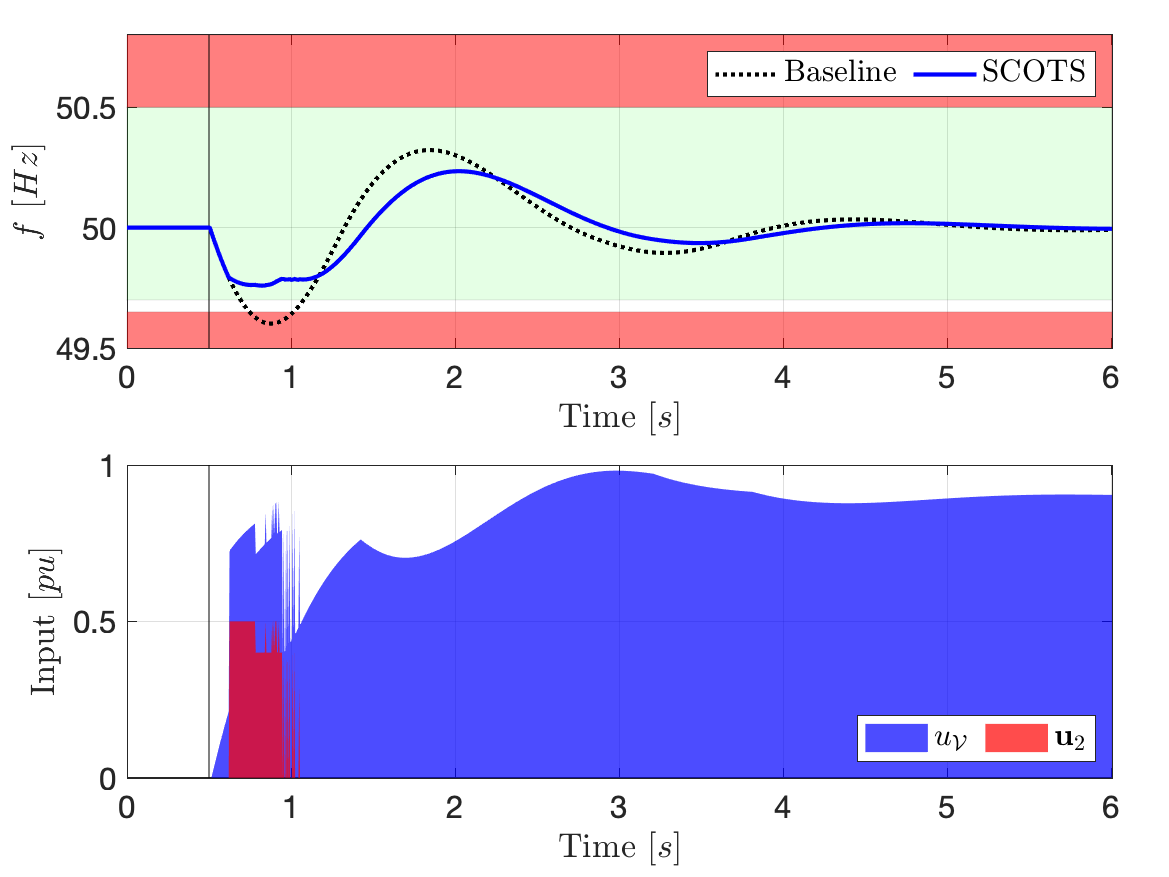}
    \centering
    \caption{\textbf{Top:} Target range $\mathcal{T}$ is shown in green, and unsafe regions $\mathcal{A}_{ub}$ and $\mathcal{A}_{lb}$ are shown in red.
    The controller designed using SCOTS and the RSF with disturbance refinement successfully satisfy $\psi$, compared with the baseline controller which violates the specification.  \textbf{Bottom:} The control input $\u_2$ designed using SCOTS for $\Sigma_2$ and the refined control input $\u_1$ for $\Sigma_1$ using our RSF.}
    \vspace{0.5cm}
    \label{fig:results}
\end{figure}

A comparison between the baseline controller and the synthesized one is provided in Figure~\ref{fig:results}.  The input $\u_2$, synthesized by SCOTS, is chosen to be the minimum $\u_2$ that guarantees satisfaction of the specification $\psi$. Successful synthesis of the controller over $\Sigma_2$ formally shows that $\psi$ also holds on $\Sigma_1$. Figure~\ref{fig:results} (bottom) shows that over the time interval $t\in [0.5,1]$, the synthesized controller over $\Sigma_2$ takes non-zero values to bring back the frequency to the intended target region, thus enabling $\Sigma_1$ to satisfy $\psi$.

As it can be observed, we provided formal guarantees using symbolic control over a $9$-dimensional system while only requiring the computational load of a $3$-dimensional system.
To verify Theorem~\ref{thm:two_final}, we quantify the maximum mismatch between output trajectories of $\Sigma_1$ and $\Sigma_2$ from simulations as \[\max_{t} \lVert \y_1(t) - \y_2(t)\rVert = 0.0541.\] Since this value is less than $\epsilon$, the controller is demonstrably formally robust.

\section{System and Specification Interconnection}
\label{sec:decomposition}
In the previous section, we showed, through a proof of concept, how to mitigate the \emph{curse of dimensionality} using simulation functions with disturbance refinement with a reduced-order model. In the proposed approach, we considered a system model which was computationally intractable to synthesise and reduced it down to a system model with lower dimensions. The trade-off was that we required a buffer $\epsilon$, which is included in the synthesis to make sure the controller is robust to the loss of information due to the reduction. The more the system is reduced, the larger $\epsilon$ will be and the greater the challenge of synthesising a robust controller. 

This approach is conservative when dealing with large-scale systems. To improve this technique for high-dimensional systems, the RSF with disturbance refinement can be combined with compositionality techniques from the literature, particularly assume-guarantee contracts. In particular, we consider the large-scale system as an interconnected network composed of several smaller subsystems. We now work on subsystems by constructing a reduced-order model abstraction for each subsystem. Under assume-guarantee contracts, we lift the results from subsystems to the interconnected system by providing formal guarantees on the satisfaction of the overall specification over the interconnected system.

\begin{figure}[h]
\vspace{0.5cm}
    \centering
    \begin{tikzpicture}[scale=1.3]
\node[draw, circle] (Sigma1) at (0,2) {$\Sigma^1$};
\node[draw, circle] (Sigma2) at (3,6) {$\Sigma^2$};
\node[draw, circle] (Sigma3) at (6,2) {$\Sigma^3$};

\draw[<-] (Sigma1) to[bend right=20] node[below,fill=white] {$\y^2_{1_2}$} (Sigma2);
\draw[<-] (Sigma2) to[bend right=20] node[below,fill=white] {$\y^1_{1_2}$} (Sigma1);
\draw[<-] (Sigma2) to[bend right=20] node[above,fill=white] {$\y^3_{1_2}$} (Sigma3);
\draw[<-] (Sigma3) to[bend right=20] node[above,fill=white] {$\y^2_{1_2}$} (Sigma2);
\draw[<-] (Sigma3) to[bend right=20] node[above,fill=white] {$\y^1_{1_2}$} (Sigma1);
\draw[<-] (Sigma1) to[bend right=20] node[below,fill=white] {$\y^3_{1_2}$} (Sigma3);
\draw[<-] (Sigma1) -- +(0,1) node[left] {$\u^1_1$};
\draw[<-] (Sigma2) -- +(-1,0) node[left] {$\u^2_1$};
\draw[<-] (Sigma3) -- +(0,1) node[right] {$\u^3_1$};
\draw[<-] (Sigma1) -- +(0,-1) node[right] {$\v^1_1$};
\draw[<-] (Sigma2) -- +(1,0) node[right] {$\v^2_1$};
\draw[<-] (Sigma3) -- +(0,-1) node[right] {$\v^3_1$};
\end{tikzpicture}
    \caption{A graphical representation of NETS composed of $3$ subsystems as vertices and interconnections with neighbours as edges.}
    \vspace{0.5cm}
    \label{fig:tikzNE39}
\end{figure}
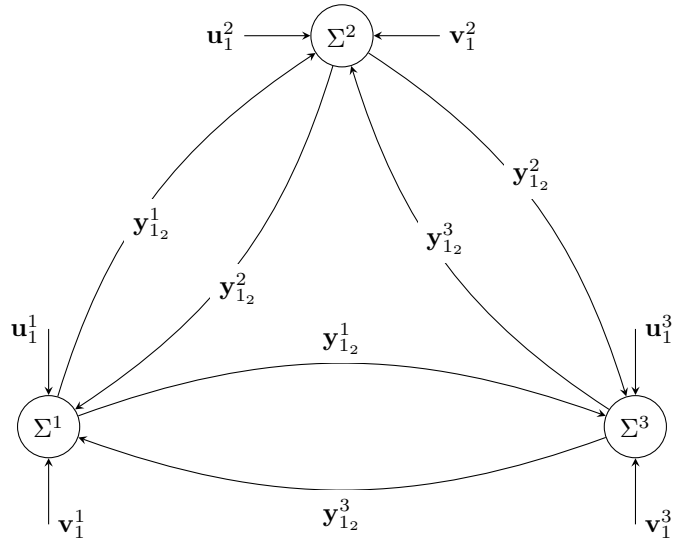

\noindent\textbf{{Running case study (continued).}}
Here, we consider NETS to be a composition of three areas, connected via the interface function on the frequency in~\eqref{eq:freq-interface}. A graphical representation of interconnections of NETS is provided in Figure~\ref{fig:tikzNE39}, in which each subsystem is labelled with its input, disturbance, and frequency. 

In the next subsection, we describe how local formal specifications of each subsystem can be combined to give a formal specification globally.

\subsection{Specification Composition}
The particular LTL properties useful for composition are conjunction ($\wedge$) and disjunction ($\vee$). Two separate LTL specifications can be combined to create either a stricter or looser specification. Therefore, in our case study, we can define local specifications on subsystems and combine them with conjunction to form a global LTL specification for the overall NETS.

\noindent\textbf{{Running case study (continued).}}
In NETS, the most important requirement is safety (or invariance), where globally the frequency must never fall beneath the containment zone. Additionally, the reachability specifications can be complementary as all subsystems are considered to have similar frequencies from the principle of interconnected systems. Therefore, all subsystems should simultaneously move toward their target area, if those targets are in a similar location.

We can write an LTL specification $\psi$ for the global system  using local specifications $\psi^i$ for all subsystems $i$ as
\begin{align*}
    &\psi^i = \LTLalways [f^i \geq \mathcal{Z}], \quad \forall i\in \{1,\ldots,N\}, \\
    &\psi = \bigwedge_{i=1}^N \psi^i.
\end{align*}
Since subsystems use~\eqref{eq:freq-interface} to transfer power between the networks, this needs to be considered in local specifications. Concisely, when synthesising controllers to guarantee the specification for a local area, the worst-case scenarios of its neighbours' actions should be considered. We do this in two different ways in the next sections: the first is to consider that the frequency in all areas is always the same (\emph{i.e.,} isolated subsystems) or to include the frequency of the other areas as an internal disturbance to the local area (\emph{i.e.,} compositionality with internal disturbances).

\section{Assume Guarantee Contracts}
\label{sec:agc}
Under assume-guarantee contracts, borrowed from~\cite{benveniste2018contracts}, we  show how subsystems can be controlled independently and combined to provide interconnected system guarantees.

\subsection{Assumptions, Guarantees and Contracts}
The properties expected from a system are called its \emph{guarantees}. Each guarantee $\mathcal{G}$ relies on a set $\alpha$ of properties called \emph{assumptions}, expressing boundary conditions for the guarantee $\mathcal{G}$ to hold.
Guarantees can be combined using conjunction, where one or more guarantees provide a contract $\mathcal{C}$. Assumptions if false remove all guarantees that relied on that assumption - other guarantees may still hold. Mathematically,
$\alpha \implies \mathcal{G}.$

For an interconnected system with multiple subsystems; assumptions and guarantees may have some independence from one another. To subsystems they may appear to be distinct subcontracts, but for the interconnected system they combine to provide strong guarantees.
So, a contract for an interconnected system with three subsystems can be defined using subcontracts
\begin{equation*}
    \mathcal{C}_1 \oplus \mathcal{C}_2 \oplus \mathcal{C}_3 \preccurlyeq \mathcal{C},
\end{equation*}
where $\oplus$ is the contract composition and $\preccurlyeq$ is a \emph{refinement relation}. For 
the general contracts $\mathcal{C}_i,~\mathcal{C}_j$ under assumptions $\alpha_i,~\alpha_j$ and providing guarantees $\mathcal{G}_i,~\mathcal{G}_j$, respectively, $\mathcal{C}_i$ refines $\mathcal{C}_j$ or $\mathcal{C}_i \preccurlyeq \mathcal{C}_j$ if and only if $\alpha_j\subseteq\alpha_i$ and $\mathcal{G}_i\subseteq\mathcal{G}_j$. 

Satisfaction of the contract is acquired when individual subsystems hold this refinement. Controllers designed on these subsystems then provide a decentralised approach to acquiring guarantees. 

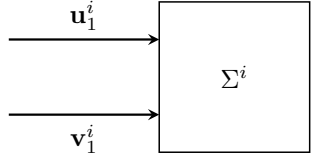
\begin{figure}[h]
\vspace{0.5cm}
    \centering
\begin{tikzpicture}[scale=2]

\node[draw, minimum width=2cm, minimum height=2cm, align=center,  line width=0.2mm] (R1) at (0,0) {$\Sigma^i$};

\draw[->,  line width=0.3mm] (-1.5,0.25) -- node[above,align=center] {$\u^i_1$} (-0.5,0.25);
\draw[->,  line width=0.3mm]  (-1.5,-0.25) -- node[below,align=center] {$\v^i_1$}(-0.5,-0.25);

\end{tikzpicture}
    \caption{Design of NETS using isolated subsystems from the principle of interconnected synchronous machines, where $i \in \{1,2,3\}$.}
    \label{fig:TikZAGC}
\end{figure}

\section{Isolated Controllers}
\label{sec:isolatedControllers}
\noindent\textbf{{Running case study (continued).}}
Here, we isolate subsystems from each other using the assumption that the frequency of the NETS is the same for all subsystems, \emph{i.e.,} $f^1 = f^2 = f^3$. This assumption follows the principle of interconnected synchronous machines. Under this assumption, the internal disturbance from neighbouring subsystems is removed, and accordingly,~\eqref{eq:freq-interface} equates to zero. An updated visual depiction of NETS is provided in Figure~\ref{fig:TikZAGC}.

We synthesize controllers to satisfy the specification (or contract) of the interconnected system using contract composition of each subsystem. If the frequencies of each area remain close to one another and a controller is designed to satisfy the specification, then the contract is satisfied for the subsystem. When all subsystem contracts are satisfied and a refinement relation holds for the contract composition, then the specification/contract for the interconnected power network is satisfied.

\subsubsection{Decentralised NETS Control}

Using a similar technique to Section~\ref{sec:PoC} with the linear case ($E^i_1=F^i_1=0$ in~\eqref{eq:nonlinear_both_disturbed_systems}),
 each area's controller can be synthesised independently with no internal disturbances between neighbouring areas due to their isolation. By combining the guarantees that the controllers provide on each area, we can ensure the global specification on NETS. Under the assumption that the area frequencies remain close to one another, each subsystem can be disturbed independently by some $\v^i_1$, and no area should violate it's specification.

\begin{figure}[h]
\vspace{0.5cm}
    \includegraphics[width=0.6\textwidth]{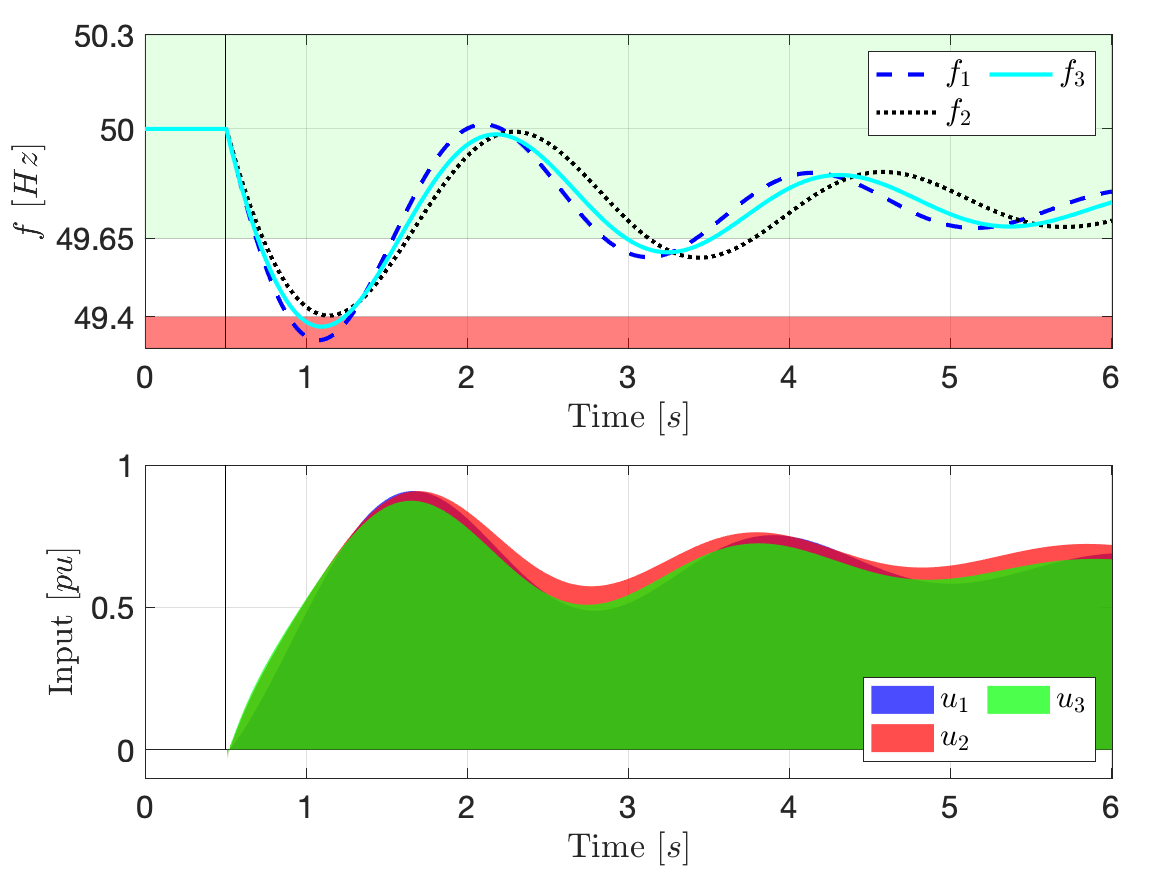}
    \centering
    \caption{\emph{Interconnected system without control.} \textbf{Top:} Target region $\mathcal{T}$ and unsafe region $\mathcal{A}$ are shown in green and red, respectively. For the baseline controller which keeps $\Sigma_1$ and $\Sigma_2$ $\epsilon$-close, it can be seen that $f^1$ falls into the red unsafe region. \textbf{Bottom:} The inputs are a byproduct of the simulation relation interface keeping $\Sigma_1$ and $\Sigma_2$ $\epsilon$-close. No additional controller is synthesised over $\Sigma_2$.}
    \vspace{0.5cm}
    \label{fig:AGC_AllO}
\end{figure}

We define the global specification as
\begin{equation*}
    \psi = \psi^1 \wedge \psi^2 \wedge \psi^3,
\end{equation*}
where $\psi^i$ is the specification of subsystem $i$. Isolating the areas also uncouples the reachability specifications providing a higher likelihood of formal guarantees over the interconnected system. For the safety guarantees, if a system can guarantee $\Delta f^i$ never falls to $-0.35~Hz$, then it implicitly guarantees $\Delta f^i$ never falls to $-0.6~Hz$. So the overall guarantee provided for all areas would match the weakest guarantee of a single area (\emph{i.e.,} worst-case scenario). In Figure~\ref{fig:AGC_AllO}, subsystems are disturbed by each $\v^i_1 = 1$ per unit ($\u^i_2 = 0$) and the frequency of multiple areas violates the safety specification. 

\begin{figure}[h!]
    \includegraphics[width=0.6\textwidth]{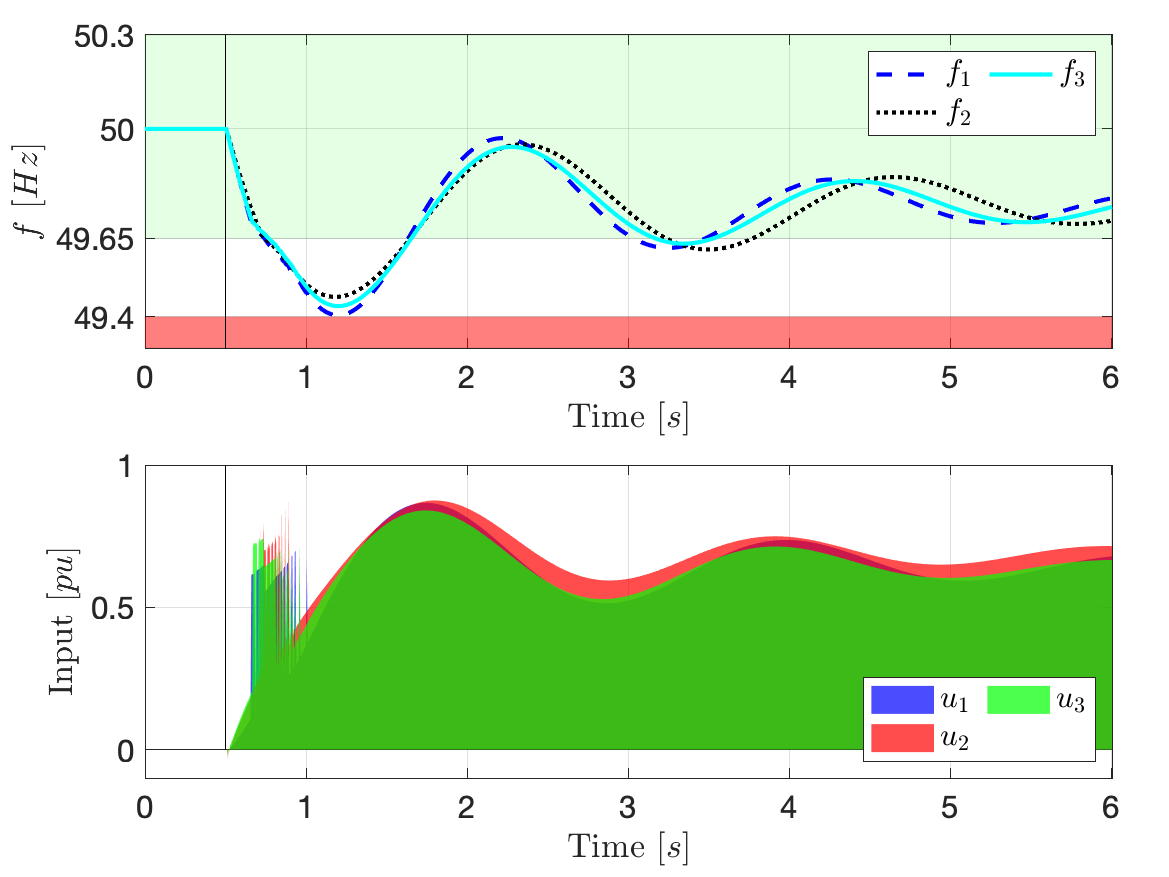}
    \centering
    \caption{\emph{Interconnected system with formal control.} \textbf{Top:} Target region $\mathcal{T}$ and unsafe region $\mathcal{A}$ are shown in green and red, respectively.
    The controller, designed using SCOTS, and the RSF with disturbance refinement successfully satisfy $\psi$.  \textbf{Bottom:} The synthesised control input for $\Sigma_2$ and the refined control input for $\Sigma_1$ using RSF are combined to provide inputs which guarantees the satisfaction of specification.}
    \vspace{0.5cm}
    \label{fig:AGC_AllC}
\end{figure}

By deploying the formal synthesised  controllers where $\u^i_2=0.5$, the assume-guarantee contract approach shows all subsystems satisfy both the safety guarantees of $\psi$ and also the reachability guarantees. The frequency of each area also remains close to the neighbouring regions, as seen in Figure~\ref{fig:AGC_AllC}.

\begin{figure}[h!]
    \includegraphics[width=0.6\textwidth]{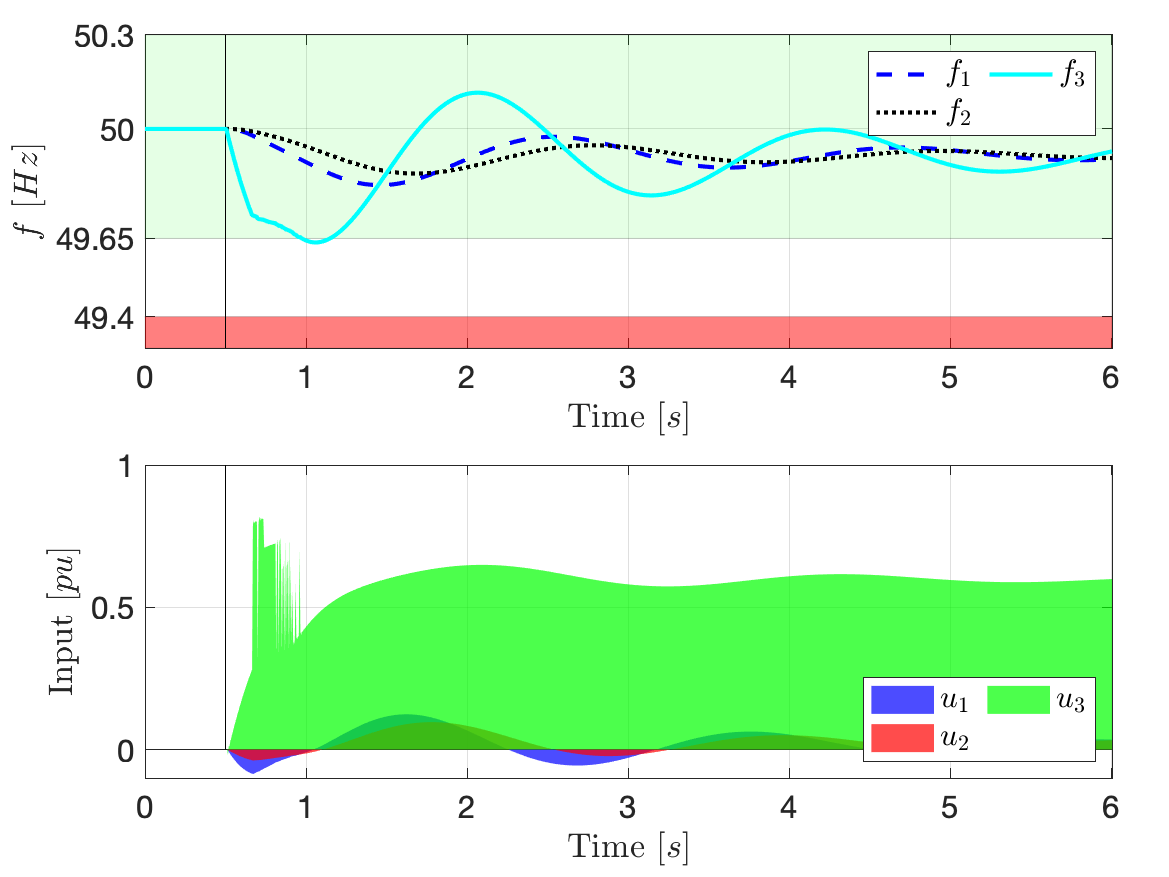}
    \centering
    \caption{\emph{Area 3 with formal control.} \textbf{Top:} Target region $\mathcal{T}$ and unsafe region $\mathcal{A}$ are shown in green and red, respectively.
    The controller, designed using SCOTS, and the RSF with disturbance refinement successfully satisfy $\psi$.  \textbf{Bottom:} The synthesised control input for $\Sigma_2$ and the refined control input for $\Sigma_1$ using RSF are combined to provide $\u^3_1$ which guarantees the satisfaction of specification.}
    \label{fig:AGC_A3C}
\end{figure}

\begin{figure}[h!]
    \includegraphics[width=0.6\textwidth]{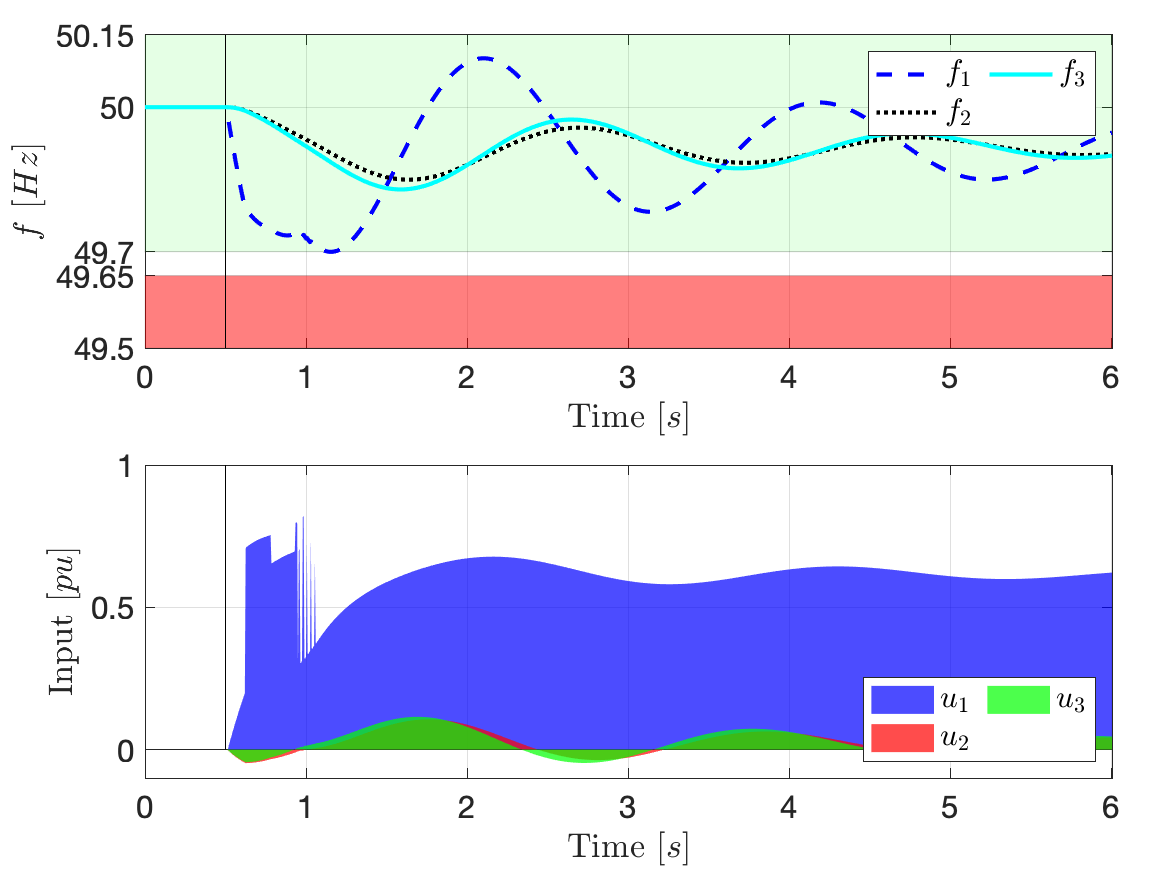}
    \centering
    \caption{\emph{Area 1 with formal control.} \textbf{Top:} Target region $\mathcal{T}$ and unsafe region $\mathcal{A}$ are shown in green and red, respectively.
    The synthesised controller and the robust simulation function with disturbance refinement successfully satisfy $\psi^1$. \textbf{Bottom:} The synthesised control input for $\Sigma_2$ and the refined control input for $\Sigma_1$ using RSF are combined to provide $\u^1_1$ which guarantees the satisfaction of specification.}
    \vspace{0.5cm}
    \label{fig:AGC_A1C}
\end{figure}

\begin{figure}[h!]
    \includegraphics[width=0.6\textwidth]{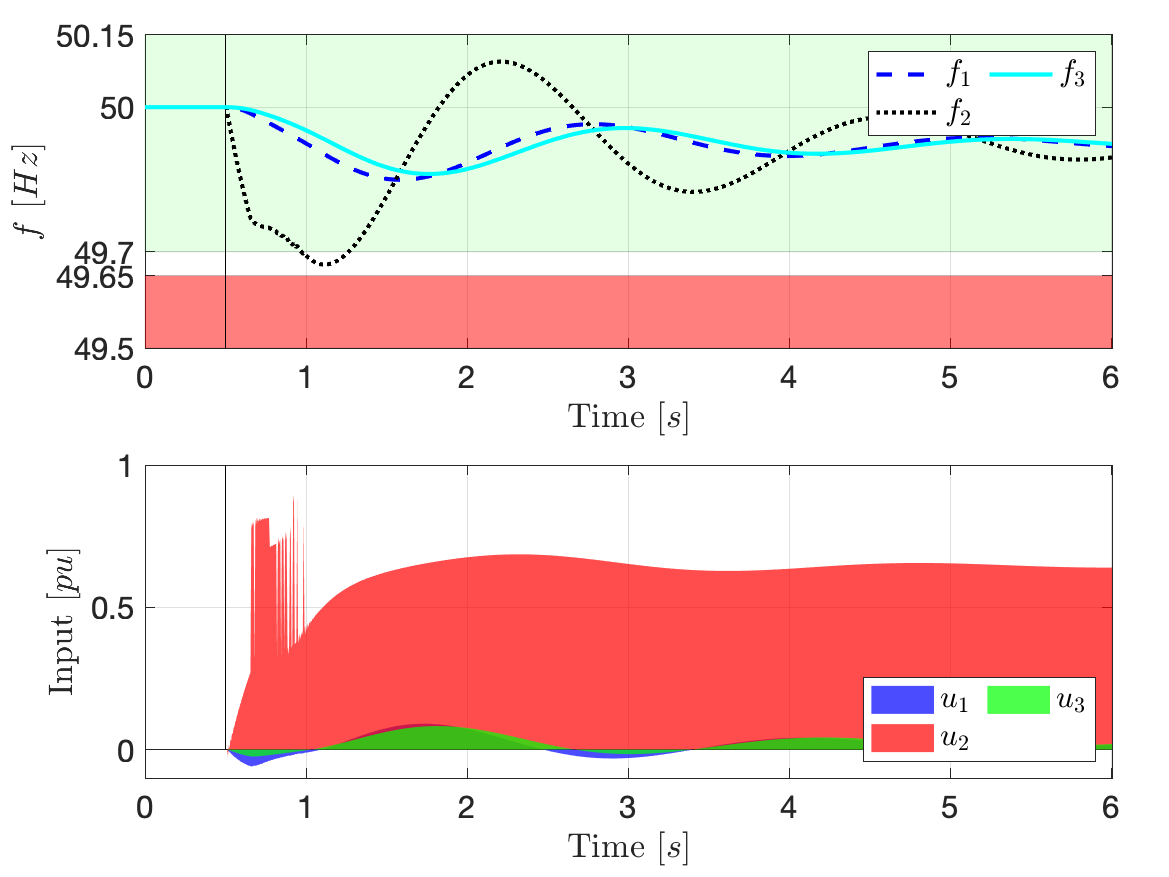}
    \centering
    \caption{\emph{Area 2 with formal control.} \textbf{Top:} Target region $\mathcal{T}$ and unsafe region $\mathcal{A}$ are shown in green and red, respectively.
    The synthesised controller and the robust simulation function with disturbance refinement successfully satisfy $\psi^2$.  \textbf{Bottom:} The synthesised control input for $\Sigma_2$ and the refined control input for $\Sigma_1$ using RSF are combined to provide $\u^2_1$ which guarantees the satisfaction of specification.}
    \label{fig:AGC_A2C}
\end{figure}

\subsubsection{Single Area Control}
We consider Area $3$ of NETS. When designing the controller, we consider a large measurable disturbance $\v^3_1$ of $1$ per unit. In addition, $\v^1_1 = \v^2_1 = 0$. $\underline{\mathcal{B}} = -0.6$, $\underline{\mathcal{T}} = -0.35$, $\overline{\mathcal{B}} = \overline{\mathcal{T}} = \infty$, $\u^3_2 = 0.5$, and $\epsilon$ is calculated as $0.1016$. The synthesised controller is depicted in Figure~\ref{fig:AGC_A3C}. It is worth remarking that for Area $3$, it was not possible to find as tight of a reach-avoid bound as in $\Sigma^1$~(Figure~\ref{fig:AGC_A1C}) and $\Sigma^2$~(Figure~\ref{fig:AGC_A2C}), where $\underline{\mathcal{B}} = -0.35$, $\underline{\mathcal{T}} = -0.3$, $\overline{\mathcal{B}} = \overline{\mathcal{T}} = \infty$.

\section{Compositionality with internal disturbances}
\label{sec:CompwII}
We now employ compositional techniques exclusively to capture internal disturbances from neighbouring subsystems. 
Under assume-guarantee contracts, we aim to strengthen the guarantees on the behaviour of the system using shared information.

\noindent\textbf{{Running case study (continued).}}
 For NETS, we consider additional knowledge of the frequency of neighbouring regions which impacts the frequency of the local subsystem. We consider the frequency of neighbouring subsystems as internal disturbances defined in~\eqref{eq:freq-interface}.
 
\begin{figure}[h]
\vspace{0.5cm}
    \centering
    \begin{tikzpicture}[scale=1.5]
\node[draw, circle] (Sigma3) at (0,2) {$\Sigma^3$};
\node[draw, circle] (Sigma1) at (2,3) {$\Sigma^1$};
\node[draw, circle] (Sigma2) at (2,1) {$\Sigma^2$};
\draw[<-] (Sigma3) to[bend right=20] node[above,fill=white] {$\y^1_{1_2}$} (Sigma1);
\draw[<-] (Sigma3) to[bend right=20] node[above,fill=white] {$\y^2_{1_2}$} (Sigma2);
\draw[<-] (Sigma3) -- +(0,1) node[left] {$\u^3_1$};
\draw[<-] (Sigma3) -- +(0,-1) node[right] {$\v^3_1$};
\end{tikzpicture}
    \caption{The NETS subsystem $\Sigma^3$ displayed as a vertex of a graph with subsystems as other vertices and interconnections with neighbours as edges. $\u^3_1$ is an input, $\v^3_1$ is an external disturbance, while $\y^1_{1_2}$ and $\y^2_{1_2}$ are internal disturbances for $\Sigma^3$.}
    \vspace{0.5cm}
    \label{fig:TikZComp}
\end{figure}
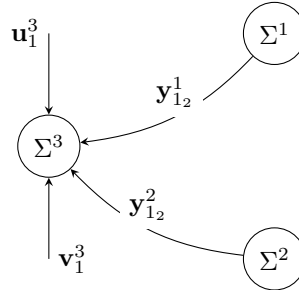

 When synthesising a controller, it is important to include the neighbouring frequency information in the synthesis procedure. Here, we consider a reach-avoid specification where each subsystem should avoid a region of the state space while trying to return a safe region after being disturbed. The disturbance of a neighbouring area should never cause the local area to violate the specification. Therefore, in the control synthesis problem, the controller should be robust to the worst-case neighbours' frequencies. Given the reach-avoid specification for each area, the boundary of the avoid region can be used to define the worst-case disturbance acting on a local subsystem from its neighbour. Figure~\ref{fig:TikZComp} shows how this looks for $\Sigma^3$ with input $\u^3_1$, external disturbance $\v^3_1$, and internal disturbances $\w^3_{1} = [\y^1_{1_2}~\y^2_{1_2}]^T$.

Using the compositional approach, subsystems have two internal disturbances coming from the neighbouring areas. For the RSF with disturbance refinement, these two disturbances have a significant impact on the value of $\epsilon$. For $\Sigma^1$, the $\epsilon$ for the isolated systems approach was $0.1016$ while for compositionality $\epsilon = 0.1896$.

\begin{figure}[h!]
    \includegraphics[width=0.6\textwidth]{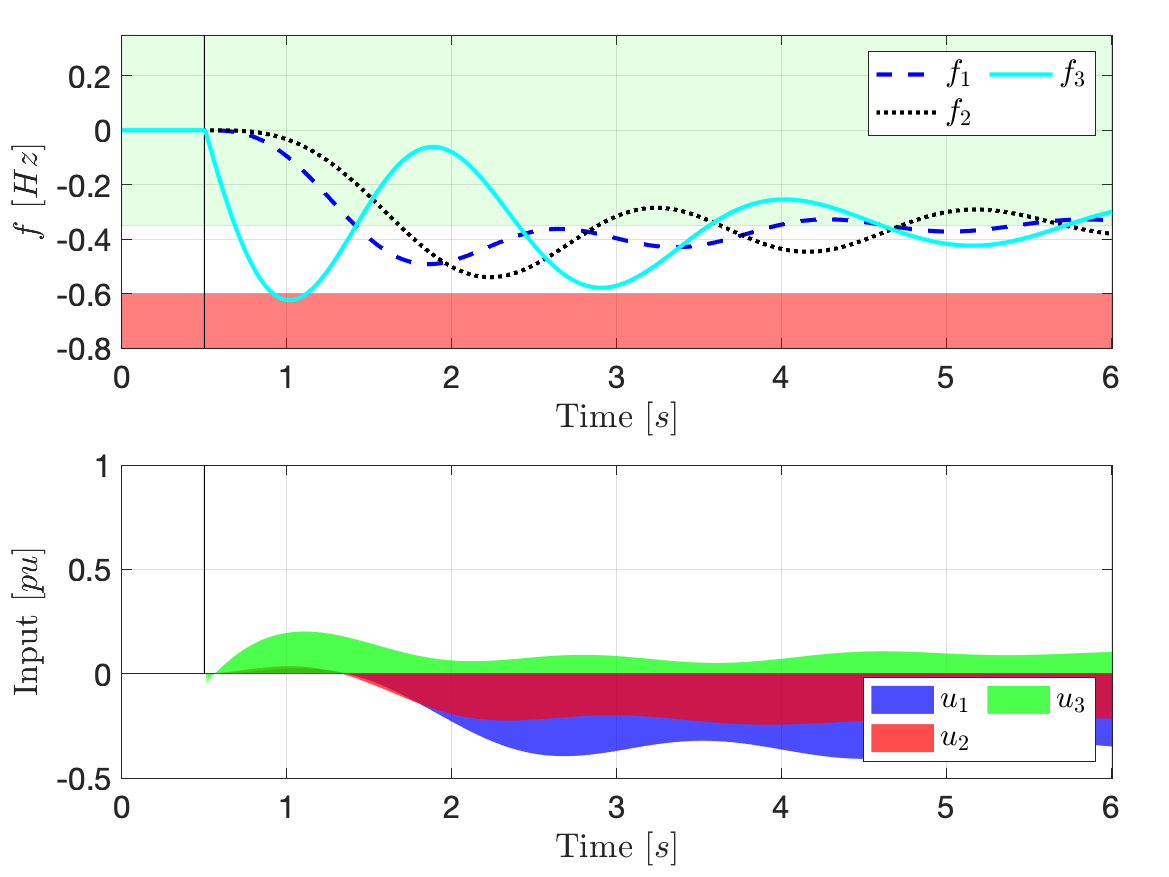}
    \centering
    \caption{\emph{Interconnected system without control.} \textbf{Top:} Target region $\mathcal{T}$ and unsafe region $\mathcal{A}$ are shown in green and red, respectively. For the baseline controller which keeps $\Sigma_1$ and $\Sigma_2$ $\epsilon$-close, it can be seen that $f^3$ falls into the red unsafe region. \textbf{Bottom:} The inputs are a byproduct of the simulation relation interface keeping $\Sigma_1$ and $\Sigma_2$ $\epsilon$-close. No additional controller is synthesised over $\Sigma_2$.
    } \vspace{0.5cm}
    \label{fig:Comp_A3O}
\end{figure}

\begin{figure}[h!]
    \includegraphics[width=0.6\textwidth]{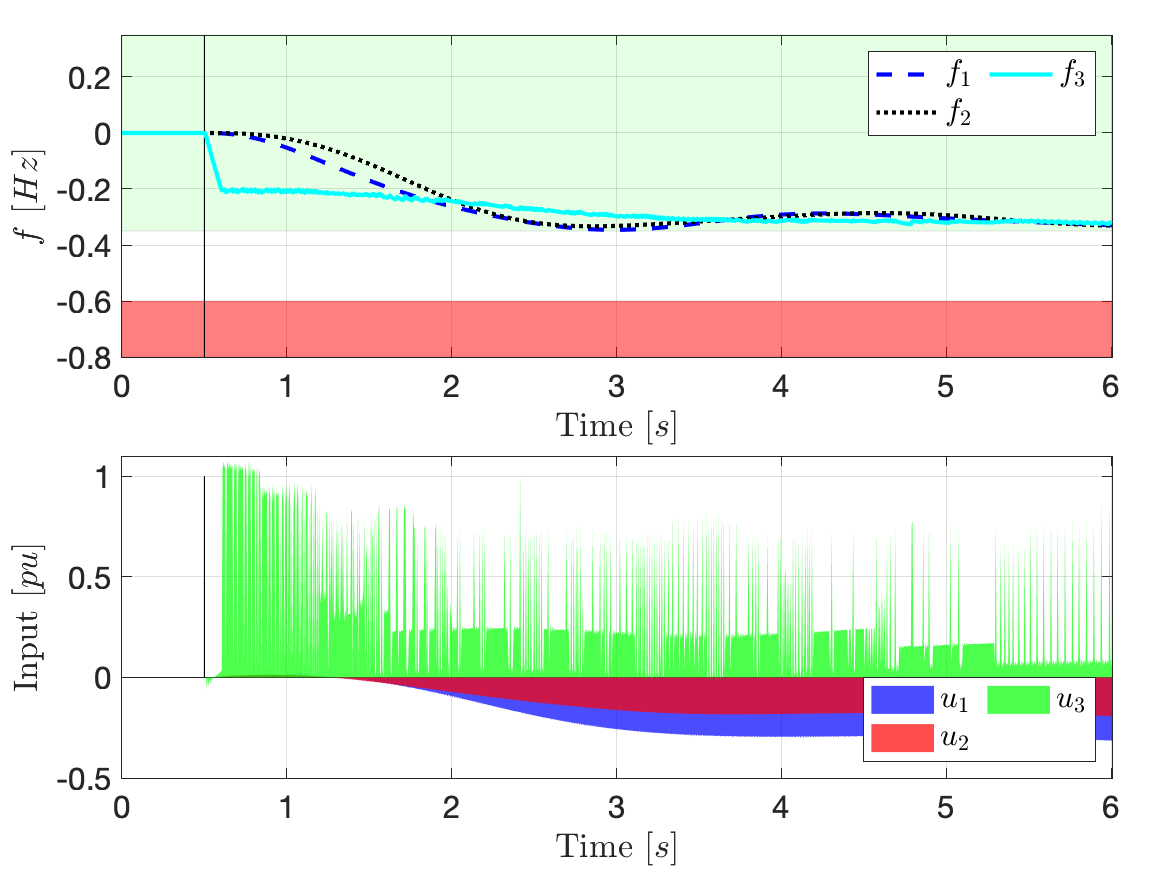}
    \centering
    \caption{\emph{Interconnected system with formal control.} \textbf{Top:} Target range $\mathcal{T}$ is shown in green, unsafe region $\mathcal{A}$ is shown in red.
    The controller designed using SCOTS and the RSF with disturbance refinement successfully satisfy $\psi$.  \textbf{Bottom:} The control input designed using SCOTS for $\Sigma_2$ and the refined control input for $\Sigma_1$ using our RSF are combined to give the inputs which guarantees the systems meets the specification.}
    \label{fig:Comp_A3C}
\end{figure}

For the compositional approach, we define a global specification that when a disturbance $\v^3_1 \leq 1$ is present in $\Sigma^3$, no areas' frequency should fall below $\Delta f^i \leq -0.6~Hz$. The simulation relation error is computed as $\epsilon^3 = 0.1992$, where $\u^3_2=1$. 
If the response of EVs is not included in the system ($\u^3_2=0$), the maximum frequency deviation violates the specification, as can be seen in Figure~\ref{fig:Comp_A3O}. Therefore, the contribution of EVs is essential to satisfy the specification on the frequency, as shown in Figure~\ref{fig:Comp_A3C}.

\section{Conclusion}
\label{sec:conclusion}
In this work, we studied a compositional control approach for large-scale power systems while providing guarantees over the system's behaviour. We employed assume-guarantee contracts with robust simulation functions (RSF) with disturbance refinement to design decentralised controllers for distinct power system areas with providing guarantees over the whole power network. We also extended the notion of RSFs with disturbance refinement to a class of \emph{nonlinear systems} and provided a temporal logic specification for frequency regulation in the GB Power Network. Throughout the paper, we used the New England $39$-Bus Test System as a challenging running case study to demonstrate our proposed approach. Future research directions can consider stochastic power systems, power system resilience and applications for voltage regulation.

\bibliographystyle{alpha}
\bibliography{biblio}

\appendix

\section{Proof of Statements}
%\vspace{-10pt}
%\noindent\textbf{Theorem 1.}
%\vspace{-10pt}
\begin{proof}[Proof of Theorem 1]
From \eqref{eq:Mhold1} and \eqref{eq:C2=C1P}, we have
\begin{align*}
\simFfull &\geq \sqrt{(\x_1-P\x_2)^TC_1^TC_1(\x_1-P\x_2)} =  \lVert C_1\x_1 - C_2\x_2\rVert,
\end{align*}
implying that condition \eqref{eq:error-bound} holds. We proceed to showing condition \eqref{eq:decay-rate}, as well. Using \eqref{eq:Mhold2} and \eqref{eq:F1=F2P}-\eqref{eq:E1=PE2}, one has
\begin{align*}
    &\frac{\partial\simF}{\partial \x_2}(A_2\x_2+B_2\u_2 +D_2d_\simF +E_2\phi(F_2\x_2)) + \frac{\partial\simF}{\partial \x_1}(A_1\x_1+B_1u_\simF+D_1\mathbf{d_1} + E_1\phi(F_1\x_1)) \\
    &\leq -\lambda\simFfull + \lVert\sqrt{M}(D_1-PD_2R_1)\d_1 + \sqrt{M}(B_1R_2-PB_2)\u_2\rVert \\
    &\leq -\lambda\simFfull + \lVert\sqrt{M}(D_1-PD_2R_1)\rVert\lVert\d_1\rVert + \lVert\sqrt{M}(B_1R_2-PB_2)\rVert\lVert\u_2\rVert.
\end{align*}
Therefore, given that
\begin{align*}
    &\frac{\lVert\sqrt{M}(D_1-PD_2R_1)\rVert}{\lambda}\lVert \d_1\rVert + \frac{\lVert\sqrt{M}(B_1R_2-PB_2)\rVert}{\lambda}\lVert\u_2\rVert \leq \simFfull,
\end{align*}
we have
\begin{align*}
    \frac{\partial\simF}{\partial \x_2}g_2(A_2\x_2+B_2\u_2 +D_2d_\simF + E_2\phi(F_2\x_2)) + \frac{\partial\simF}{\partial \x_1}g_1(A_1\x_1+B_1u_\simF+D_1\mathbf{d}_1 + E_1\phi(F_1\x_1)) \leq 0,
\end{align*}
which completes the proof.
\end{proof}
%\vspace{-10pt}
%\noindent\textbf{Theorem 2.}

\begin{proof}[Proof of Theorem 2]
For the sake of an easier presentation, we slightly abuse notation and denote $\simF(\x_1(t),\x_2(t))$ by $\simF(t)$.
Let
\begin{equation*}
    \epsilon = \max\{\simF(0), \gamma_1(\lVert\d_1\rVert_\infty) + \gamma_2(\lVert\u_2\rVert_\infty)\}.
\end{equation*} 
We show $\simF(t) \leq \epsilon$ for all $t$.
Showing $\simF(0) \leq \epsilon$ is straightforward due to the definition of $\epsilon$. We show the rest of the proof based on contradiction.
Assume there exists $\tau > 0$ such that $\simF(\tau) > \epsilon$. Then there also exists some $0 \leq \tau' < \tau$ such that $\simF(\tau') = \epsilon$ and $\forall t\in(\tau',\tau],\simF(t) > \epsilon$. Note that we have, $\forall t\in(\tau',\tau]$,
\begin{align*}
    \gamma_1(\lVert\d_1\rVert) + \gamma_2(\lVert\u_2\rVert) \leq \gamma_1(\lVert\d_1\rVert_\infty) + \gamma_2(\lVert\u_2\rVert_\infty) \leq \epsilon < \simF(t).
\end{align*}
From \eqref{eq:decay-rate}, we then have $ \frac{\partial \simF(t)}{\partial t} \leq 0$ for all $t\in(\tau',\tau]$,
which implies
\begin{equation*}
    \simF(\tau) - \simF(\tau') = \int^\tau_{\tau'}\frac{\partial \simF(t)}{\partial t}\partial t \leq 0.
\end{equation*}
This contradicts $\simF(\tau) > \epsilon = \simF(\tau')$. Therefore, $\simF(t) \leq \epsilon$ for all $t$.
Finally from \eqref{eq:error-bound}, we have:
\begin{equation*}
    \simF(\x_1(t),\x_2(t)) \leq \epsilon \implies \lVert\y_1(t)-\y_2(t)\rVert \leq \epsilon.
\end{equation*}
\end{proof}

\section{Matrices of New England 39-Bus Test System}
\noindent\textbf{Proof of Concept - Nonlinear Area 1}

   $ A_1 =  \begin{bmatrix}
    -12.5 &	0 &	0 &	0.09 &	-0.65 &	0&	0&	0&	-0.09 \\
0&	-16.67&	0&	0.09&	-0.65&	0&	0&	0&	-0.09\\
0&	0&	-14.29&	0.05&	-0.61&	0&	0&	0&	-0.05\\
0&	0&	0&0&	0.93&	0&	0&	0&	0\\
0&	0&	0&	-6.28&	-0.09&	2.5&	2.78&	2.38&	0\\
12.5&	0&	0&	0&	0&	-2.5&	0&	0	&0\\
0&	16.67&	0&	0&	0&	0&	-2.78&	0&	0\\
0&	0&	14.29&	0&	0&	0&	0&	-2.38&	0\\
0&	0&	0&	6.28&	2.08&	0&	0&	0&	0\\
    \end{bmatrix}\\
    B_1 = \begin{bmatrix}
    0 &0&0&0&1&0&0&0&0
    \end{bmatrix}^T \\ 
    D_1 = \begin{bmatrix}
    0 &0&0&0&-1&0&0&0&0
    \end{bmatrix}^T  \\
 C_1 = \begin{bmatrix}
    0&	0&	0&	0&	2.05&	0&	0&	0&	0
    \end{bmatrix}\\
     E_1 = \begin{bmatrix}
    0 &0&0&0&0.0285&0&0&0&0
    \end{bmatrix}^T \\
     F_1 = \begin{bmatrix}
    0&	0&	0&	0&	2&	0&	0&	0&	0
    \end{bmatrix} \\$
$
    A_2 =  \begin{bmatrix}
    -0.6333 &	3.0028 &	0.4428 \\
    -3.0028 &	-0.0026&	-0.0263\\
    -0.4428&	-0.0263&	-1.5159\\
    \end{bmatrix}\\
    B_2 = \begin{bmatrix}
    -1.0204&0.6395&0.8273
    \end{bmatrix}^T \\ 
    D_2 = \begin{bmatrix}
    1.0204&-0.6395&-0.8273
    \end{bmatrix}^T  \\
    C_2 = \begin{bmatrix}
    -1.5128&	0.096&	0.5044
    \end{bmatrix}\\
     E_2 = \begin{bmatrix}
    1.0204&-0.6395&-0.8273
    \end{bmatrix}^T \\ 
     F_2 = \begin{bmatrix}
    -1.4726& 0.0934&	0.4910
    \end{bmatrix} \\
    M =  \begin{bmatrix}
0.22 &	0 &	0 &	0 &	0.01 &	-0.01&	0&	0&	0 \\
0&	0.26&	0&	0&	0.01&	0&	-0.01&	0&	0\\
0&	0&	0.26&	0&	0.01&	0&	0&	-0.01&	0\\
0&	0&	0&  82.14&	20.22&	0&	0&	0&	16.80\\
0.01&	0.01&	0.01&	20.22&	11.62&	0&	0&	0&	11.68\\
-0.01&	0&	0&	0.01&	0&	0.02&	0&	0	&0\\
0&	-0.01&	0&	0.01&	0&	0&	0.02&	0&	0\\
0&	0&	-0.01&	0.01&	0&	0&	0&	0.02&	0\\
0&	0&	0&	16.80&	11.68&	0&	0&	0&	29.44\\
    \end{bmatrix}\\
    P = \begin{bmatrix}
    0.03&0.025&0.03&0.01&-0.74&0.02&0.04&0.025&0.55\\
    -0.01&-0.01&-0.01&0.25&0.05&-0.08&-0.08&-0.08&0.3\\
    -0.02&-0.015&-0.015&-0.15&0.25&-0.28&-0.21&-0.26&0.44\\
    \end{bmatrix}^T \\ 
    Q_1 = K_1 = \begin{bmatrix}
    0&0&0&0&0&0&0&0&0
    \end{bmatrix}  \\
    K_2 = \begin{bmatrix}
    -0.2&-0.2&-0.2&-482.5&-278.9&-2.5&-2.8&-2.4&-279.9
    \end{bmatrix}\\
     Q_2 = \begin{bmatrix}
    0.02&-0.0343&0.2860
    \end{bmatrix}\\
    R_1=R_2=1,
    L_{11} = L_{21} = 0, L_{12} = 0.0285, L_{22} = 1,
    ESS_{max} = 0.5454, ESS_{min} = 0, \overline{\delta} = 1.$

\noindent\textbf{Fully Interconnected New England 39-Bus System}

\begin{equation*}
 B =\left[\scalemath{1}{
    \begin{array}{ccccccccccccccccccccccccccc}
0 & 0 & 0 & 0 & 1 & 0 & 0 & 0 & 0 & 0 & 0 & 0 & 0 & 0 & 0 & 0 & 0 & 0 & 0 & 0 & 0 & 0 & 0 & 0 & 0 & 0 & 0 \\
0 & 0 & 0 & 0 & 0 & 0 & 0 & 0 & 0 & 0 & 0 & 0 & 1 & 0 & 0 & 0 & 0 & 0 & 0 & 0 & 0 & 0 & 0 & 0 & 0 & 0 & 0 \\
0 & 0 & 0 & 0 & 0 & 0 & 0 & 0 & 0 & 0 & 0 & 0 & 0 & 0 & 0 & 0 & 0 & 0 & 0 & 0 & 1 & 0 & 0 & 0 & 0 & 0 & 0
\end{array}
}\right]^T
\end{equation*}
\begin{equation*}
 C =\left[\scalemath{1}{
    \begin{array}{ccccccccccccccccccccccccccc}
0 & 0 & 0 & 0 & 2.05 & 0 & 0 & 0 & 0 & 0 & 0 & 0 & 0 & 0 & 0 & 0 & 0 & 0 & 0 & 0 & 0 & 0 & 0 & 0 & 0 & 0 & 0 \\
0 & 0 & 0 & 0 & 0 & 0 & 0 & 0 & 0 & 0 & 0 & 0 & 1.83 & 0 & 0 & 0 & 0 & 0 & 0 & 0 & 0 & 0 & 0 & 0 & 0 & 0 & 0 \\
0 & 0 & 0 & 0 & 0 & 0 & 0 & 0 & 0 & 0 & 0 & 0 & 0 & 0 & 0 & 0 & 0 & 0 & 0 & 0 & 2.09 & 0 & 0 & 0 & 0 & 0 & 0
\end{array}
}\right]
\end{equation*}

\begin{equation*}
 D =\left[\scalemath{1}{
    \begin{array}{ccccccccccccccccccccccccccc}
0 & 0 & 0 & 0 & -1 & 0 & 0 & 0 & 0 & 0 & 0 & 0 & 0 & 0 & 0 & 0 & 0 & 0 & 0 & 0 & 0 & 0 & 0 & 0 & 0 & 0 & 0 \\
0 & 0 & 0 & 0 & 0 & 0 & 0 & 0 & 0 & 0 & 0 & 0 & -1 & 0 & 0 & 0 & 0 & 0 & 0 & 0 & 0 & 0 & 0 & 0 & 0 & 0 & 0 \\
0 & 0 & 0 & 0 & 0 & 0 & 0 & 0 & 0 & 0 & 0 & 0 & 0 & 0 & 0 & 0 & 0 & 0 & 0 & 0 & -1 & 0 & 0 & 0 & 0 & 0 & 0
\end{array}
}\right]^T
\end{equation*}

\begin{landscape}

\begin{equation*}
 A =\left[\scalemath{0.6}{
    \begin{array}{ccccccccccccccccccccccccccc}
-12.5 &	0	&0	&0.09	&-0.65	&0	&0	&0	&0	&0	&0	&0	&0	&0	&0	&0	&0	&0	&0	&0	&0	&0	&0	&0	&-0.11	&0	&0\\
0&	-16.67&	0&	0.09&	-0.65&	0&	0&	0&	0&	0&	0&	0&	0&	0&	0&	0&	0&	0&	0&	0&	0&	0&	0&	0&	-0.11&	0&	0 \\
0&	0&	-14.29&	0.05&	-0.61&	0&	0&	0&	0&	0&	0&	0&	0&	0&	0&	0&	0&	0&	0&	0&	0&	0&	0&	0&	-0.05&	0&	0 \\
0&	0&	0&	0&	0.92&	0&	0&	0&  0&	0&	0&	0&	-0.37&	0&	0&	0&	0&	0&	0&	0&	-0.52&	0&	0&	0&	0&	0&	0 \\
0&	0&	0&	-6.28&	-0.09&	2.5&	2.78&	2.38&	0&	0&	0&	0&	0&	0&	0&	0&	0&	0&	0&	0&	0&	0&	0&	0&	0&	0&	0 \\
12.5&	0&	0&	0&	0&	-2.5&	0&	0&	0&	0&	0&	0&	0&	0&	0&	0&	0&	0&	0&	0&	0&	0&	0&	0&	0&	0&	0 \\
0&	16.67&	0&	0&	0&	0&	-2.78&	0&	0&	0&	0&	0&	0&	0&	0&	0&	0&	0&	0&	0&	0&	0&	0&	0&	0&	0&	0 \\
0&	0&	14.29&	0&	0&	0&	0&	-2.38&	0&	0&	0&	0&	0&	0&	0&	0&	0&	0&	0&	0&	0&	0&	0&	0&	0&	0&	0 \\
0&	0&	0&	0&	0&	0&	0&	0&	-16.67&	0&	0&	0.18&	-0.61&	0&	0&	0&	0&	0&	0&	0&	0&	0&	0&	0&	0&	-0.19&	0 \\
0&	0&	0&	0&	0&	0&	0&	0&	0&	-16.67&	0&	0&	-0.68&	0&	0&	0&	0&	0&	0&	0&	0&	0&	0&	0&	0&	0&	0 \\
0&	0&	0&	0&	0&	0&	0&	0&	0&	0&	-12.5&	0.12&	-0.69&	0&	0&	0&	0&	0&	0&	0&	0&	0&	0&	0&	0&	-0.13&	0 \\
0&	0&	0&	0&	-0.41&	0&	0&	0&	0&	0&	0&	0&	0.58&	0&	0&	0&	0&	0&	0&	0&	-0.25&	0&	0&	0&	0&	0&	0 \\
0&	0&	0&	0&	0&	0&	0&	0&	0&	0&	0&	-6.28&	-0.08&	2.27&	3.13&	2.5&	0&	0&	0&	0&	0&	0&	0&	0&	0&	0&	0 \\
0&	0&	0&	0&	0&	0&	0&	0&	16.67&	0&	0&	0&	0&	-2.27&	0&	0&	0&	0&	0&	0&	0&	0&	0&	0&	0&	0&	0 \\
0&	0&	0&	0&	0&	0&	0&	0&	0&	16.67&	0&	0&	0&	0&	-3.13&	0&	0&	0&	0&	0&	0&	0&	0&	0&	0&	0&	0 \\
0&	0&	0&	0&	0&	0&	0&	0&	0&	0&	12.5&	0&	0&	0&	0&	-2.5&	0&	0&	0&	0&	0&	0&	0&	0&	0&	0&	0 \\
0&	0&	0&	0&	0&	0&	0&	0&	0&	0&	0&	0&	0&	0&	0&	0&	-14.29&	0&	0&	0&	-0.74&	0&	0&	0&	0&	0&	0 \\
0&	0&	0&	0&	0&	0&	0&	0&	0&	0&	0&	0&	0&	0&	0&	0&	0&	-14.29&	0&	0.12&	-0.65&	0&	0&	0&	0&	0&	-0.17 \\
0&	0&	0&	0&	0&	0&	0&	0&	0&	0&	0&	0&	0&	0&	0&	0&	0&	0&	-12.5&	0.12&	-0.67&	0&	0&	0&	0&	0&	-0.17 \\
0&	0&	0&	0&	-0.51&	0&	0&	0&	0&	0&	0&	0&	-0.22&	0&	0&	0&	0&	0&	0&	0&	0.77&	0&	0&	0&	0&	0&	0 \\
0&	0&	0&	0&	0&	0&	0&	0&	0&	0&	0&	0&	0&	0&	0&	0&	0&	0&	0&	-6.28&	-0.096&	3.33&	2.5&	2.44&	0&	0&	0 \\
0&	0&	0&	0&	0&	0&	0&	0&	0&	0&	0&	0&	0&	0&	0&	0&	14.29&	0&	0&	0&	0&	-3.33&	0&	0&	0&	0&	0 \\
0&	0&	0&	0&	0&	0&	0&	0&	0&	0&	0&	0&	0&	0&	0&	0&	0&	14.29&	0&	0&	0&	0&	-2.5&	0&	0&	0&	0 \\
0&	0&	0&	0&	0&	0&	0&	0&	0&	0&	0&	0&	0&	0&	0&	0&	0&	0&	12.5&	0&	0&	0&	0&	-2.44&	0&	0&	0 \\
0&	0&	0&	6.28&	2.08&	0&	0&	0&	0&	0&	0&	0&	0&	0&	0&	0&	0&	0&	0&	0&	0&	0&	0&	0&	-1&	0&	0 \\
0&	0&	0&	0&	0&	0&	0&	0&	0&	0&	0&	6.28&	2.16&	0&	0&	0&	0&	0&	0&	0&	0&	0&	0&	0&	0&	-1&	0 \\
0&	0&	0&	0&	0&	0&	0&	0&	0&	0&	0&	0&	0&	0&	0&	0&	0&	0&	0&	6.28&	2.24&	0&	0&	0&	0&	0&	-1 \\
\end{array}
}\right]
\end{equation*}
\end{landscape}

\noindent\textbf{Linear Area 1 - Isolated}

$ A_1 =  \begin{bmatrix}
    -12.5 &	0 &	0 &	0.09 &	-0.65 &	0&	0&	0&	-0.09 \\
0&	-16.67&	0&	0.09&	-0.65&	0&	0&	0&	-0.09\\
0&	0&	-14.29&	0.05&	-0.61&	0&	0&	0&	-0.05\\
0&	0&	0&0&	0.93&	0&	0&	0&	0\\
0&	0&	0&	-6.28&	-0.09&	2.5&	2.78&	2.38&	0\\
12.5&	0&	0&	0&	0&	-2.5&	0&	0	&0\\
0&	16.67&	0&	0&	0&	0&	-2.78&	0&	0\\
0&	0&	14.29&	0&	0&	0&	0&	-2.38&	0\\
0&	0&	0&	6.28&	2.08&	0&	0&	0&	0\\
    \end{bmatrix}\\
    B_1 = \begin{bmatrix}
    0 &0&0&0&1&0&0&0&0
    \end{bmatrix}^T \\ 
    D_1 = \begin{bmatrix}
    0 &0&0&0&-1&0&0&0&0
    \end{bmatrix}^T  \\
    C_1 = \begin{bmatrix}
    0&	0&	0&	0&	2.05&	0&	0&	0&	0
    \end{bmatrix}\\
    A_2 =  \begin{bmatrix}
    -0.6333 &	3.0028 &	0.4428 \\
    -3.0028 &	-0.0026&	-0.0263\\
    -0.4428&	-0.0263&	-1.5159\\
    \end{bmatrix}\\
    B_2 = \begin{bmatrix}
    -0.8580&0.5378&0.6956
    \end{bmatrix}^T \\ 
    D_2 = \begin{bmatrix}
    0.8580&-0.5378&-0.6956
    \end{bmatrix}^T  \\
    C_2 = \begin{bmatrix}
    -1.7990&	0.1141&	0.5998
    \end{bmatrix} $\\
   $ M =  \begin{bmatrix}
0.22 &	0 &	0 &	0 &	0.01 &	-0.01&	0&	0&	0 \\
0&	0.26&	0&	0&	0.01&	0&	-0.01&	0&	0\\
0&	0&	0.26&	0&	0.01&	0&	0&	-0.01&	0\\
0&	0&	0&  82.14&	20.22&	0&	0&	0&	16.80\\
0.01&	0.01&	0.01&	20.22&	11.62&	0&	0&	0&	11.68\\
-0.01&	0&	0&	0.01&	0&	0.02&	0&	0	&0\\
0&	-0.01&	0&	0.01&	0&	0&	0.02&	0&	0\\
0&	0&	-0.01&	0.01&	0&	0&	0&	0.02&	0\\
0&	0&	0&	16.80&	11.68&	0&	0&	0&	29.44\\
    \end{bmatrix}\\
    P = \begin{bmatrix}
    0.04&0.03&0.03&0.02&-0.88&0.025&0.044&0.03&0.66\\
    -0.01&-0.01&-0.01&0.29&0.06&-0.10&-0.98&-0.99&0.36\\
    -0.03&-0.18&-0.018&-0.18&0.29&-0.33&-0.25&-0.31&0.52\\
    \end{bmatrix}^T \\ 
    Q_1 = K_1 = \begin{bmatrix}
    0&0&0&0&0&0&0&0&0
    \end{bmatrix}  \\
    K_2 = \begin{bmatrix}
    -0.2&-0.2&-0.2&-482.5&-278.9&-2.5&-2.8&-2.4&-279.9
    \end{bmatrix}\\
     Q_2 = \begin{bmatrix}
    0.0238&-0.0407&0.3401
    \end{bmatrix}\\
    R_1=R_2=1.$

Area 1 is formed from rows and columns $1-8$ and $25$ from $A$ in the fully interconnected system. 

\noindent\textbf{Linear Area 1 - with Internal Disturbance}

The internal disturbances can be derived from column $13$ and column $21$ of $A$ returning the following disturbance matrix.

$D_1 = \begin{bmatrix}
    0 &0&0&0&-1&0&0&0&0 \\
    0 &0&0&-0.37&0&0&0&0&0 \\
    0 &0&0&-0.52&0&0&0&0&0 \\
    \end{bmatrix}^T$  \\

\noindent\textbf{Linear Area 2 and Area 3}

Area 2 and Area 3 can be derived in the same way that Area 1 was derived using the fully interconnected New England 39-Bus System. The matrices for these areas are not provided.
\end{document}